\documentclass[10pt,twocolumn,letterpaper]{article}

\usepackage{iccv}              

\usepackage{lipsum}
\usepackage[ruled,vlined]{algorithm2e} 
\usepackage{calc}
\usepackage{microtype}
\usepackage{graphicx}
\usepackage[accsupp]{axessibility}
\usepackage{xcolor}
\usepackage{booktabs} 
\usepackage{colortbl}  
\usepackage{xcolor}
\usepackage{multirow}

\usepackage{amsmath}
\usepackage{amssymb}
\usepackage{mathtools}
\usepackage{nopageno}
\usepackage{amsthm}
\usepackage{pifont}
\renewcommand{\thefootnote}{}

\definecolor{iccvblue}{rgb}{0.21,0.49,0.74}
\usepackage[pagebackref,breaklinks,colorlinks,allcolors=iccvblue]{hyperref}

\def\paperID{4251} 
\def\confName{ICCV}
\def\confYear{2025}

\author{
    Yusheng Dai$^{1*}$, Chenxi Wang$^{1*}$, Chang  Li$^1$,
    Chen Wang$^2$, Kewei Li$^1$, Jun Du$^{1\dag}$, \\ Lei Sun$^3$, Jianqing Gao$^3$, Ruoyu Wang$^1$, Jiefeng Ma$^1$ \\
    $^1$ University of Science and Technology of China, Hefei, China \\
     $^2$ Tsinghua University, Beijing, China $^3$ iFlytek Research, Hefei, China 
}

\title{Latent Swap Joint Diffusion for 2D Long-Form Latent Generation}

\begin{document}
\maketitle
\footnotetext{ * Equal Contribution. $^{\dag}$ Corresponding author.}
\setcounter{footnote}{0}
\renewcommand{\thefootnote}{\arabic{footnote}}

\begin{abstract}

\vspace{-12pt} 
This paper introduces Swap Forward (SaFa), a modality-agnostic and efficient method to generate seamless and coherent long spectrum and panorama using a latent swap joint diffusion process across multi-views. We first investigate spectrum aliasing problem in spectrum-based audio generation caused by existing joint diffusion methods. Through a comparative analysis of the VAE latent representation of spectra and RGB images, we identify that the failure arises from excessive suppression of high-frequency components due to the step-wise averaging operator. To address this issue, we propose Self-Loop Latent Swap, a frame-level bidirectional swap operator, applied to the overlapping region of adjacent views. Leveraging step-wise differentiated trajectories, this swap operator avoids spectrum distortion and adaptively enhances high-frequency components. Furthermore, to improve global cross-view consistency in non-overlapping regions, we introduce Reference-Guided Latent Swap, a unidirectional latent swap operator that provides a centralized reference trajectory to synchronize subview diffusions. By refining swap timing and intervals, we canachieve a balance between cross-view similarity and diversity in a feed-forward manner. Quantitative and qualitative experiments demonstrate that SaFa significantly outperforms existing joint diffusion methods and even training-based methods in audio generation using both U-Net and DiT models. It also adapts well to panorama generation, achieving comparable performance with a 2 $\times$ to 20 $\times$ speedup. The project website is available at \href{https://swapforward.github.io}{https://swapforward.github.io}.

\end{abstract}
\setlength{\parindent}{1.5em}
\vspace{-15pt}
\section{Introduction}
\vspace{-5pt}
\label{sec:introduction}

\begin{table*}[!t]
\vspace{-5pt}
\centering
\small
\setlength{\tabcolsep}{3.5mm}
\begin{tabular}{lcccccc}
\toprule[1pt]
\textbf{Method}    & \textbf{\begin{tabular}[c]{@{}c@{}}Spectrum\\ Adaptation\end{tabular}} & \textbf{\begin{tabular}[c]{@{}c@{}}Feed\\ Forward\end{tabular}} & \textbf{\begin{tabular}[c]{@{}c@{}}Fixed Attention\\ Window\end{tabular}}  & \textbf{\begin{tabular}[c]{@{}c@{}}DiT\\ Adaptation\end{tabular}} & \textbf{\begin{tabular}[c]{@{}c@{}}Guidance\\ Method\end{tabular}} & \textbf{Runtime (s)}  \\
\hline
MultiDiffusion \cite{BarTal2023MultiDiffusionFD}     & \text{\ding{55}}      & \text{\ding{52}}           & \text{\ding{52}}      & \text{\ding{52}}    & -                & 37.71            \\
MAD \cite{Quattrini2024MergingAS}            & \text{\ding{55}}      & \text{\ding{52}}          & \text{\ding{55}}      & \text{\ding{55}}    & Self-Attention   & 41.82           \\
SyncDiffusion  \cite{lee2023syncdiffusion} & \text{\ding{55}}      & \text{\ding{55}}         & \text{\ding{52}}      & \text{\ding{52}}    & Gradient Descent & 390.63        \\
\textbf{Swap Forward} & \text{\ding{52}}      & \text{\ding{52}}        & \text{\ding{52}}       & \text{\ding{52}}     & Latent Swap   & \textbf{19.15}           \\           
\bottomrule[1pt]

\end{tabular}
\vspace{-5pt}

\caption{Swap Forward demonstrates strong \textit{generalization} (adapting to both spectrum and image generation, U-Net and DiT architectures, and fixed or flexible attention window sizes), \textit{simplicity} (applying only two latent swap operators in a feed-forward manner), and \textit{efficiency} ($2 \sim 20\times$ faster inference time with less subview count) to generate seamless and coherent long-form audio and panorama.}
\vspace{-15pt}

\label{tab:com}
\end{table*}

Diffusion models learn to generate data by progressively adding noise to existing samples and subsequently reversing this process to recover the original data \cite{ho2020denoising, song2020score}. They initially achieved remarkable success in text-to-image \cite{ho2020denoising, song2020score, song2020denoising} and rapidly expanded into text-to-video \cite{ he2022latent, wang2023gen, qiu2023freenoise, kim2024fifo} and text-to-audio \cite{Liu2023AudioLDMTG, ghosal2023text, Huang2023MakeAnAudio2T, Evans2024FastTL,  Huang2023Noise2MusicTM, li2024quality, liu2024audioldm}. Despite these advances, modeling the variability of the physical world remains challenging, particularly when diffusion models are trained on domain-specific datasets. Therefore, expanding pretrained diffusion models to generate a broader range of data types has become increasingly attractive. One significant challenge is length extrapolation, which aims to generate images of arbitrary shapes or audio of varying lengths using diffusion models trained on data with fixed sizes.


For text-to-image generation, panorama generation can be approached as a length extrapolation problem, involving extensive pixel or latent sequences with extreme aspect ratios. Related training-free methods generally fall into two categories: autoregressive approaches \cite{lu2024autoregressive, Avrahami2022BlendedLD, Kim2024FIFODiffusionGI, liu2025panofree} and joint diffusion approaches \cite{jimenez2023mixture, Quattrini2024MergingAS, lee2023syncdiffusion, zhang2023diffcollage, BarTal2023MultiDiffusionFD}. Compared with the former, joint diffusion methods have attracted broader research interest in recent years due to their high efficiency, lower error accumulation and do not suffer from temporal causality constraints. Generally, related work mainly focuses on two issues: achieving smooth transitions between adjacent subviews and maintaining cross-view consistency, e.g., color and style, in distant subviews.

As representative work \cite{jimenez2023mixture, BarTal2023MultiDiffusionFD}, MultiDiffusion (MD) achieves smooth transitions between adjacent views by averaging their noisy latent maps at each denoising step. In this case, a new joint diffusion process is optimized by synchronizing several subview diffusion processes to produce a globally coherent output. However, due to a lack of explicit guidance, this approach relies on a high overlap rate that suffering from low efficiency, but still leads to global perceptual mismatches  across subviews. As an advanced version of MD, Merge-Attention-Diffuse (MAD) \cite{Quattrini2024MergingAS} merges and splits subview latent maps around the self-attention layer, enabling global attention over the entire panorama. However, the merge operation increases the self-attention window length, forcing the model to handle longer token sequences and introduce position embedding repetition problem. Thus, it is observed to lead to significant performance degradation in DiT architectures in Tab.~\ref{tab:qual} and in spectrum generation models trained on short-length clips in Tab.~\ref{tab:audio1}. SyncDiffusion \cite{lee2023syncdiffusion} proposes a promising guidance approach to enhance the global coherence by minimizing perceptual similarity (LPIPS loss) \cite{zhang2018unreasonable} between each subview and the reference image. However, the additional forward and backward propagation can significantly increase computation and time costs. Meanwile, LPIPS loss is observed to be insensitive to \textit{intermediate} denoised mel-spectrograms. \textit{Consequently, there remains a lack of efficient and architecture-agnostic joint diffusion methods that can achieve global cross-view consistent in both panorama and audio generation.}

For audio generation, long-form soundscapes and background music are in high demand for ambiance enhancement in real-life applications (e.g., in-car audio, sleep aids) and digital products (e.g., movies, video games). Most related works focus on training-based long-form audio generation (AudioGen \cite{kreuk2022audiogen} and Stable audio \cite{evans2024long}), which incurs significant training costs and is sensitive to text prompts. Few studies have explored the applicability of joint diffusion methods in spectrum-based audio generation. In early experiment, we try to apply existing joint diffusion methods to spectrum-based audio generation \cite{polyak2024movie}. However, as shown in Fig.~\ref{fig:head} and Fig.~\ref{fig:qual}, we observe a spectrum aliasing phenomenon: \textit{the generated spectrograms exhibit low time-frequency resolution and distortion in overlap transition regions (narrow white bands), leading to visual blurriness and low-quality audio with distortion and monotonous tails.} This phenomenon is particularly evident in spectrally rich audio including more spectral details, e.g., soundscapes and concertos.

To address these issues, in Section \ref{sec:Spectrum}, we first investigate above spectrum aliasing phenomenon in spectrum generation caused by existing joint diffusion. Specifically, leveraging the connectivity inherent to constant, we propose, we provide a comparative analysis of the VAE latent representations across the Mel-specta and RGB images connected with their original feature, exhibiting the high-frequency variability of the spectrum latent map. Through Fourier analysis, we further uncover that the failure stems from the excessive suppression of high-frequency components in the spectrum denoising process caused by the latent averaging operator.

Based on these findings, in Section \ref{sec:latent swap}, we propose the Self-loop Latent Swap operator, a frame-level bidirectional latent swap operator applied in the overlapping regions of adjacent subviews. Relying on the step-wise differentiated trajectories of adjacent subviews, it effectively addresses the spectrum aliasing problem by adaptively enhancing spectra high-frequency details in the overlapping regions. Moreover, to achieve cross-view consistency in the non-overlapping regions, we further propose the Reference-Guided Latent Swap, a unidirectional latent swap operator that provides a centralized reference trajectory for each subview diffusion in early denoising steps. By adjusting the timing and interval of the swap operation, we can balance the similarity-diversity trade-off across subviews in a feed-forward manner. \textit{Through these two latent swap operators, SaFa not only fills the gap for existing joint diffusion methods in long spectrum generation, but also acts as a highly efficient alternative for generating cross-view consistent panoramas with significantly reduced time cost and subview number.}

Finally, in Section \ref{sec:exp}, we present extensive quantitative and qualitative experiments on both long-form audio and panorama generation tasks with U-Net and DiT diffusion models. Compared to state-of-the-art (SOTA) joint diffusion methods and even training-based approaches, SaFa demonstrates much greater simplicity and efficiency, achieving superior generation quality through only two fundamental swap operators, ensuring smooth transitions and cross-view consistency with significantly reduced time cost. As shown in Tab.~\ref{tab:com}, we further provide a detailed key point comparison of SaFa and existing joint diffusion methods.

\vspace{-3pt}
\section{Related Works}
\paragraph{Long-Form Audio Generation}
Related work focuses mainly on training-based methods, including Language Models (LMs) \cite{agostinelli2023musiclm, copet2023simple, borsos2023audiolm} and Diffusion Models (DMs) \cite{evans2024long, Evans2024FastTL, tan2024litefocus}. LMs, typically based on auto-regressive architectures, face temporal causality constraints \cite{agostinelli2023musiclm, copet2023simple, borsos2023audiolm, kreuk2022audiogen}, leading to increasing accumulated errors and repetition issues when generating long audio. DMs mainly focus on 10s durations \cite{liu2023audioldm, ghosal2023tango, majumder2024tango, Huang2023MakeAnAudio2T}, and while Make-An-Audio2 \cite{Huang2023MakeAnAudio2T} supports variable-length generation, it struggles with longer durations. Stable Audio \cite{Evans2024FastTL, evans2024long} is trained in long-form audio, but it incurs high training costs and is sensitive to text prompts. Meanwhile, related work on joint diffusion generation in audio generation remains limited \cite{polyak2024movie}.
\vspace{-3pt}

\paragraph{Panorama Generation}
Early training-free methods \cite{Avrahami2022BlendedLD, Avrahami2021BlendedDF, esser2021taming} apply mainly painting techniques with DM, which easily cause repetition problems and suffer from temporal causality constraints \cite{lee2023syncdiffusion, Quattrini2024MergingAS}. Recent advancements in panorama generation mainly based on joint diffusion, as mentioned in Introduction. Furthermore, SyncTweedies \cite{kim2024synctweedies} and StochSync \cite{yeo2025stochsync} extend joint diffusion to 3D mesh texturing through Tweedie’s formula and Score Distillation Sampling, respectively. Meanwhile, we note a class of training-free methods, such as ScaleCrafter and DemoFusion \cite{he2023scalecrafter, du2024demofusion}, mainly applied to high-resolution image generation (e.g., upscaling portraits), which can also be formulated as a 2D length extrapolation problem. These methods focus on upscaling at the pixel level, while ensuring global structural coherence while preventing object repetition. In contrast, this paper focuses on long-spectrum (e.g., concertos, soundscapes) and panorama generation (e.g., mountains, crowds), emphasizing smooth subview transitions, cross-view similarity-diversity balance with extending more objects (e.g., pitches, harmonics).

\vspace{-3pt}

\section{Preliminary}

\begin{figure*}[!t]
    \centering
\includegraphics[width=2.1\columnwidth]{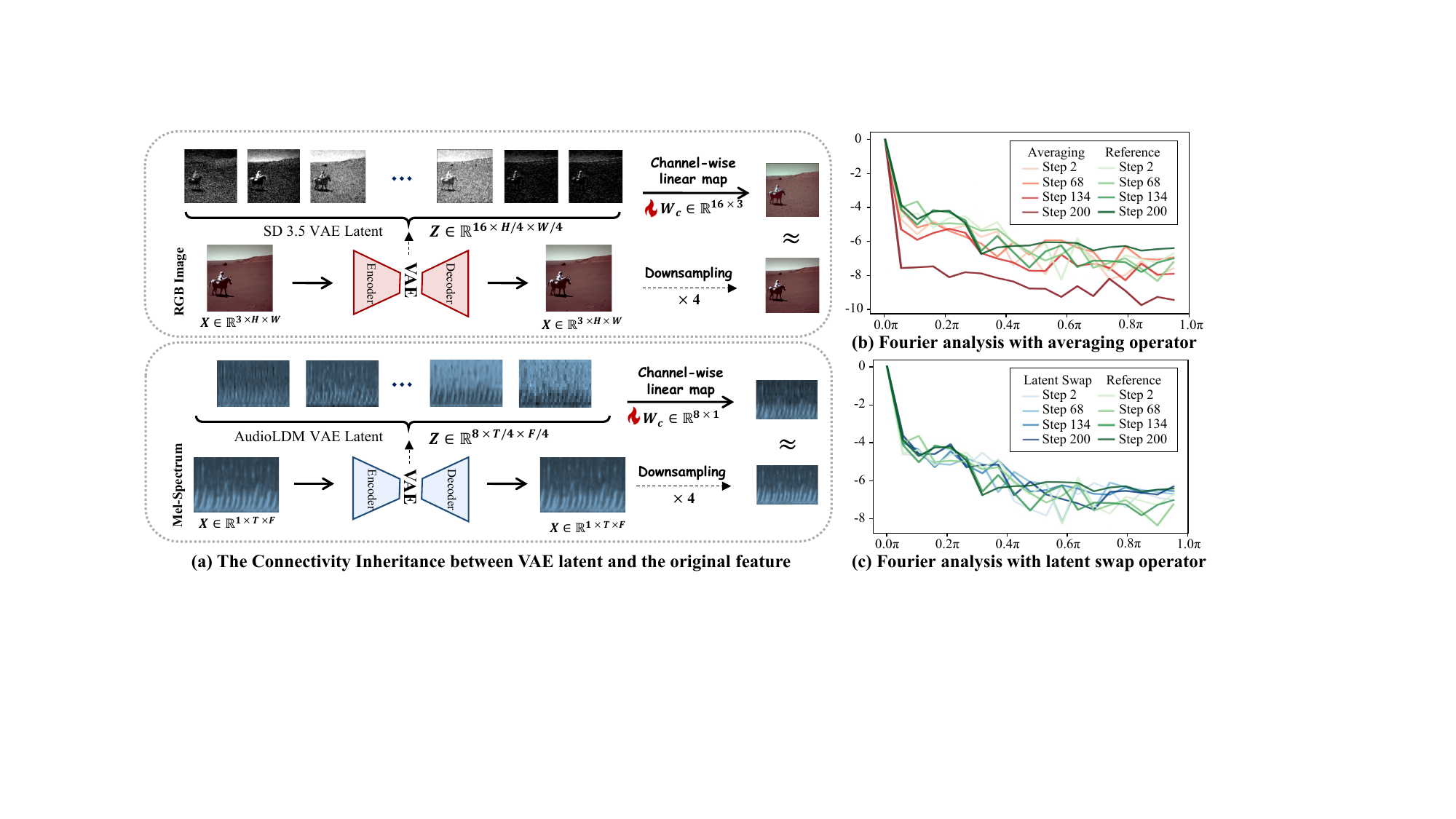}
 \vspace{-15pt}
    \caption{(a) The channel-wise linear approximation between original feature downsampling and its VAE latent ensures their Connectivity Inheritance, maintaining connectivity and structure consistency. (b) (c) Relative log-amplitude curves in 2D Fourier analysis for non-overlapping regions (Reference) and overlapping regions of the denoised spectrum latent map with the averaging and latent swap operators, where  X and Y axes represent frequency and relative log amplitude.}
\label{analysis}
\vspace{-15pt}
\end{figure*}

\begin{figure*}[!t]
    \centering
\includegraphics[width=2.1\columnwidth]{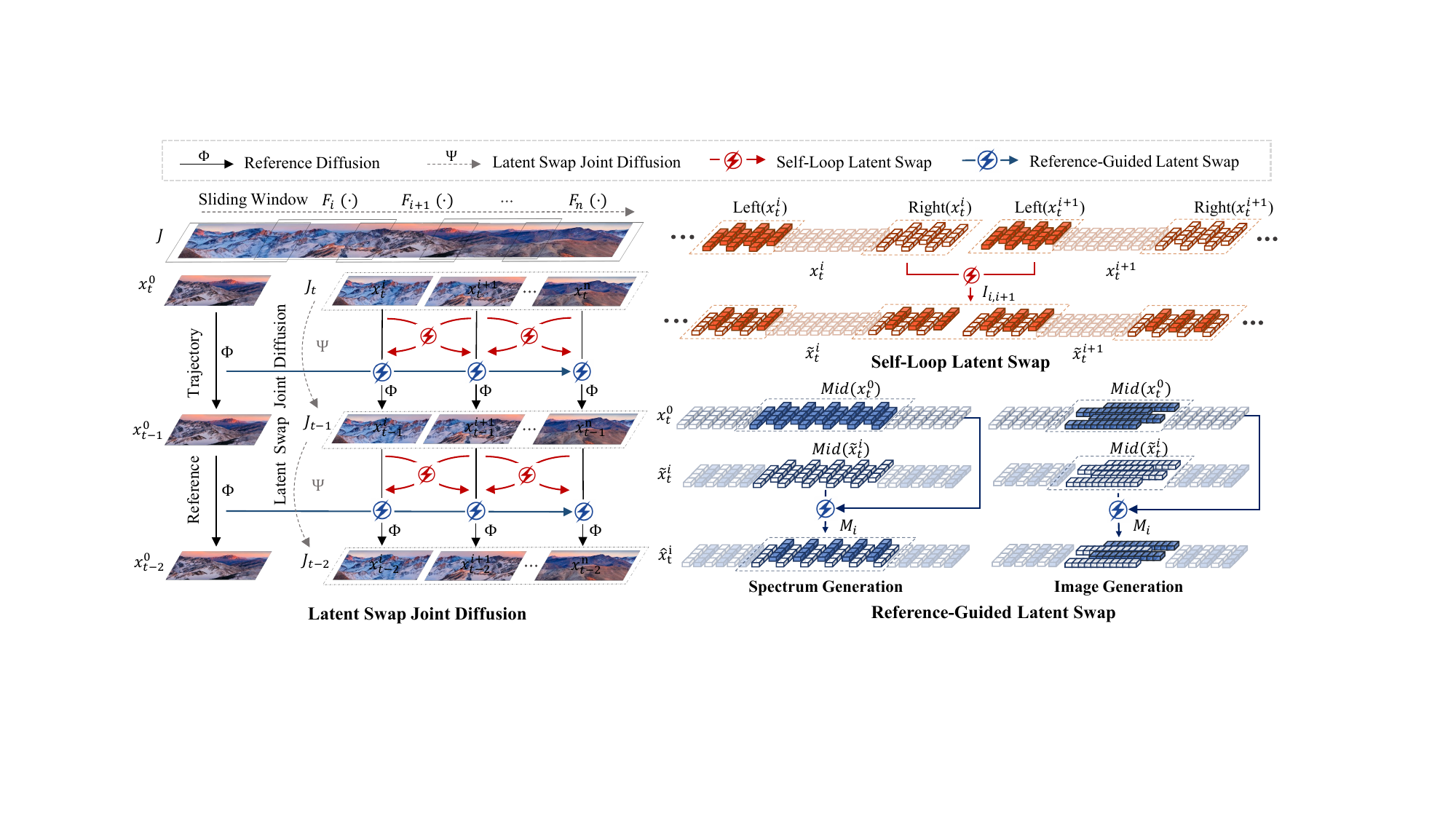}
\vspace{-15pt}
    \caption{Our latent swap joint diffusion pipeline. As the core, the Self-Loop Latent Swap operator performs bidirectional frame-level swaps on the overlapping regions of adjacent subviews during each denoising step, adaptively enhancing the high-frequency details and avoiding spectrum aliasing. The Reference-Guided Latent Swap occurs between the reference and each subview trajectories during the early steps, providing a centralized reference trajectory to ensure cross-view consistency without repetition.}
\label{fig:middle}
\vspace{-15pt}
\end{figure*}

\paragraph{Joint Diffusion for Long Latent Generation}
Diffusion models initially operate in the original feature space \cite{ho2020denoising} and have recently been extended to the VAE latent space, achieving higher fidelity and compression rates \cite{rombach2022high}. For most modality generation tasks, they can be reformulated as a general Latent Diffusion process in the corresponding VAE latent space. As show in the Fig.~\ref{fig:middle} left, given a reference latent diffusion model $\Phi$ and conditions $\{y_i\}^n_{i=0}$, our target is to generate a 2D long-form latent map $ J \in \mathbb{R}^{C \times H \times W}$ ($H \ll W$) through a joint diffusion process $\Psi$ by merging a sequence of subview latent maps $\{X^i\}_{i=1}^n \in \mathbb{R}^{C \times H_x \times W_x}$. The subview mapping $F_i$ maps the overview $J$ to subview $X_i$ as:
\begin{equation}
F_i : J \rightarrow X_i , i \in [n]
\end{equation}
which can be considered as a 1D sliding window process. Considering that commonly in the spectrum $H \ll W$, we simplify the overlap mapping $I_{i, i+1}$ from overall $J$ to the overlap region of the adjacent subviews $X_i$ and $X_{i+1}$  as:
\begin{equation}
I_{i, i+1} \leftrightarrow F_i \cap F_{i+1}
\label{I}
\end{equation}
and the non-overlap mapping $M_{i}$ of subview $X_i$ as:
\begin{equation}
    M_{i} \leftrightarrow F_i - \left(F_{i-1} \cup F_{i+1}\right)
\end{equation}
For joint diffusion step $\Psi\left(J_t \mid Y\right)$ at step $t$ on condition $Y$, most work \cite{BarTal2023MultiDiffusionFD,lee2023syncdiffusion, jimenez2023mixture} apply averaging operator to synchronize different denoising trajectories in overlap regions,
\begin{equation}
\small
\Psi\left(J_t \mid Y\right)=\sum_{i=1}^n \frac{F_i^{-1}\left(W_i\right)}{\sum_{j=1}^n F_j^{-1}\left(W_j\right)} \odot F_i^{-1}\left(\Phi(X_t^i \mid y_i)\right)
\label{trajectory}
\end{equation}
where $W_i$ is the weight matrix of subview $X_i$ that is implemented with averaging in overlapping regions and exclusivity in nonoverlapping regions. $F^{-1}$ is reversed mapping of $F$ from J to $X_i$ by zero padding.

\vspace{-3pt}
\section{Investigation on VAE Latent}
\label{sec:Method}
\label{sec:Spectrum}
To address the spectrum aliasing problem in overlap regions as shown in Fig.~\ref{fig:head}, we first conduct a comparative analysis of the spectrum and image VAE representation, which received limited prior investigation. Initially, our main insight stems from the differences between the original features. The mel-spectrogram is a 2D time-frequency representation of the audio signal, showing amplitude variability across frequency bands over time. Compared to RGB images, mel-spectrograms exhibit high-frequency variability, with amplitude distributions in time-frequency bins showing weak connectivity, marked by sparsity and discreteness, lacking the continuous contours seen in normal images. 

Further, in Fig.~\ref{analysis} (a), we observe that this discrepancy can extend to the VAE latent space due to the Connectivity Inheritance between the original feature and its VAE latent. And the inheritance is governed by a learnable, global,  channel-wise linear approximation mapping $W_c$, which is specific to each VAE model. Specifically, given an image or spectrum $X \in \mathbb{R}^{C_x \times W_x \times H_x}$ and its VAE latent representation $Z \in \mathbb{R}^{C_z \times W_z \times H_z}$, we define a learnable constant linear mapping $W_c \in \mathbb{R}^{C_x \times C_z}$ along the channel dimension , satisfying the following approximation:
\begin{equation}
\text{Downsample}(X) \approx W_c \cdot Z
\end{equation}
Correspondingly, the inverse mapping can be implemented as:
\begin{equation}
Z \approx (W_c^T W_c)^{-1} W_c^T \cdot \text{Downsample}(X)
\end{equation}
Such linear approximation mapping along the channel dimension ensures connectivity and structural alignment between the latent map and the original features. Consequently, as shown in Fig.~\ref{analysis} (a)  the spectrum latent consistently exhibits low connectivity and high-frequency variability compared to image latent. So far, we have validated the Connectivity Inheritance of VAE latents across most convolution-based VAEs used in image generation (SD 2.0 \cite{rombach2022high} and SD 3.5 \cite{esser2024scaling}) and audio generation ( AudioLDM \cite{liu2023audioldm}, Tango \cite{ghosal2023tango} and Make-An-Audio2 \cite{Huang2023MakeAnAudio2T}), which can be attributed to the local spatial correspondence in convolution operations.


We further analyze the impact of the averaging operator on spectrum latents diffusion via Fourier analysis. As shown in Fig.~\ref{analysis}(b), unlike RGB images \cite{si2024freeu}, the relative amplitude curve of the reference non-overlapping regions spectrum exhibits dynamic fluctuations but lacks a clear reduction in high-frequency components during denoising due to its high-frequency variability. However, in overlap regions, the step-wise averaging operation progressively smooths high-frequency components, leading to a more pronounced decline, particularly in later denoising steps. This causes spectral detail loss and spectrum aliasing, manifesting as visual blurriness and auditory distortion or monotonous tails.

\section{Latent Swap Joint Diffusion}
\label{sec:latent swap}
\paragraph{Step-wise Differentiated Trajectories}
\label{sec:Trajectories}Building on the above findings, our target is to enhance high-frequency details while spectrum aliasing. As shown in black and red trajectories in Fig.~\ref{fig:middle} left, focusing on overlapping regions, the essence of joint diffusion can be regarded as the merging of step-wise differentiated trajectories $\Phi(x_t^i \cup  I_{i, i+1}(J_{t+1}),y_i)$ and  $\Phi(x_t^{i+1} \cup I_{i, i+1}(J_{t+1}),y_{i+1})$ that share the same previous-step initial overlapping latent {\small $I_{i, i+1}(J_{t+1})$}. Thus, these step-wise trajectories exhibit both differences—arising from latent influences outside the overlap and similarities—inherited from the shared initial state at the previous step, which can be formulated as:
\begin{equation}
\begin{aligned}
 \varepsilon_l \leqslant d \big(&\operatorname{Right}\big(\Phi(x_t^i \cup  I_{i, i+1}(J_{t+1}),y_i)\big)\big), \\
 &\operatorname{Left}\big(\Phi(x_t^{i+1} \cup I_{i, i+1}(J_{t+1}),y_{i+1})\big) \big) \leq \varepsilon_u 
\end{aligned}
\end{equation}
where the lower bound \( \varepsilon_l \in [0,1] \) prevents complete similarity, the upper bound \( \varepsilon_u \in [0,1] \) restricts excessive divergence. $\text{Right}(\cdot)$ and $\text{Left}(\cdot)$ stand for region slice operations on the subview to obtain the left and right overlap regions, and \( d(\cdot) \) is a distance metric.


\vspace{-10pt}
\paragraph{Self-Loop Latent Swap}
 \label{sec:Self-Loop}
Building on the properties of step-wise differentiated trajectories, in this subsection, we introduce the latent swap operator $W_\text{swap}$, which consists of binary elements 0 and 1 as a specific type of linear combination operator. First, when focusing on the latent merge operation in the overlap region, the averaging operator is commonly applied with a constant matrix with constant scalar $c$ as:
\begin{equation}
W_\text{avg} = c \cdot 1_{m \times n}
\end{equation}
In contrast, in Fig.~\ref{fig:middle} our swap joint diffusion applies a binary swap operator $W_\text{swap}$ to preserve the subcomponent of the denoised latents from the step-wise differential trajectories $x_t^i$ and $x_t^{i+1}$ rather than smoothing them together as:

\begin{equation}
\scalebox{0.83}{$I_{i,i+1}(J_t) = W_\text{swap} \odot \text{Right}(X_t^i) + (1-W_\text{swap}) \odot \text{Left}(X_t^{i+1})$}
\label{swap_equation}
\end{equation}

where $I(\cdot)$ denotes the overlap mapping operation defined in Eq.~\ref{I}. We claim that this hard-combination approach is robust due to the similarity of step-wise differentiated trajectories (originating from the same overlap latent) and the inherent stability of the diffusion model. Furthermore, the latent swap operator functions analogously to an adaptive filter based on the differential of step-wise differentiated trajectories (with independent non-overlapping regions), adaptively enhancing specific frequency components by controlling the swap interval $w$, defined as:

\begin{equation}
W_\text{swap} = \mathbf{1}_n \otimes v_m, \quad
v_m^{(i)} = \frac{1}{2} \left[ 1 - (-1)^{\left\lfloor \frac{i-1}{w} \right\rfloor} \right]
\label{swapo}
\end{equation}

Based on the experimental results in Appendix, we select an optimal swap interval of $w=1$, which serves as a frame-level latent swap operation and results in smoother, better-blended transitions. At a high level, the latent swap operation is applied to the overlapping region $I_{i,i+1}$ sequentially across each subview, including between the first and last subviews, forming a looped swap process without central guidance. thus we refer to this approach as Self-Loop Latent Swap.

\begin{figure}[!t]
    \centering
    \setlength{\belowcaptionskip}{0pt} 
    \includegraphics[width=1\columnwidth]{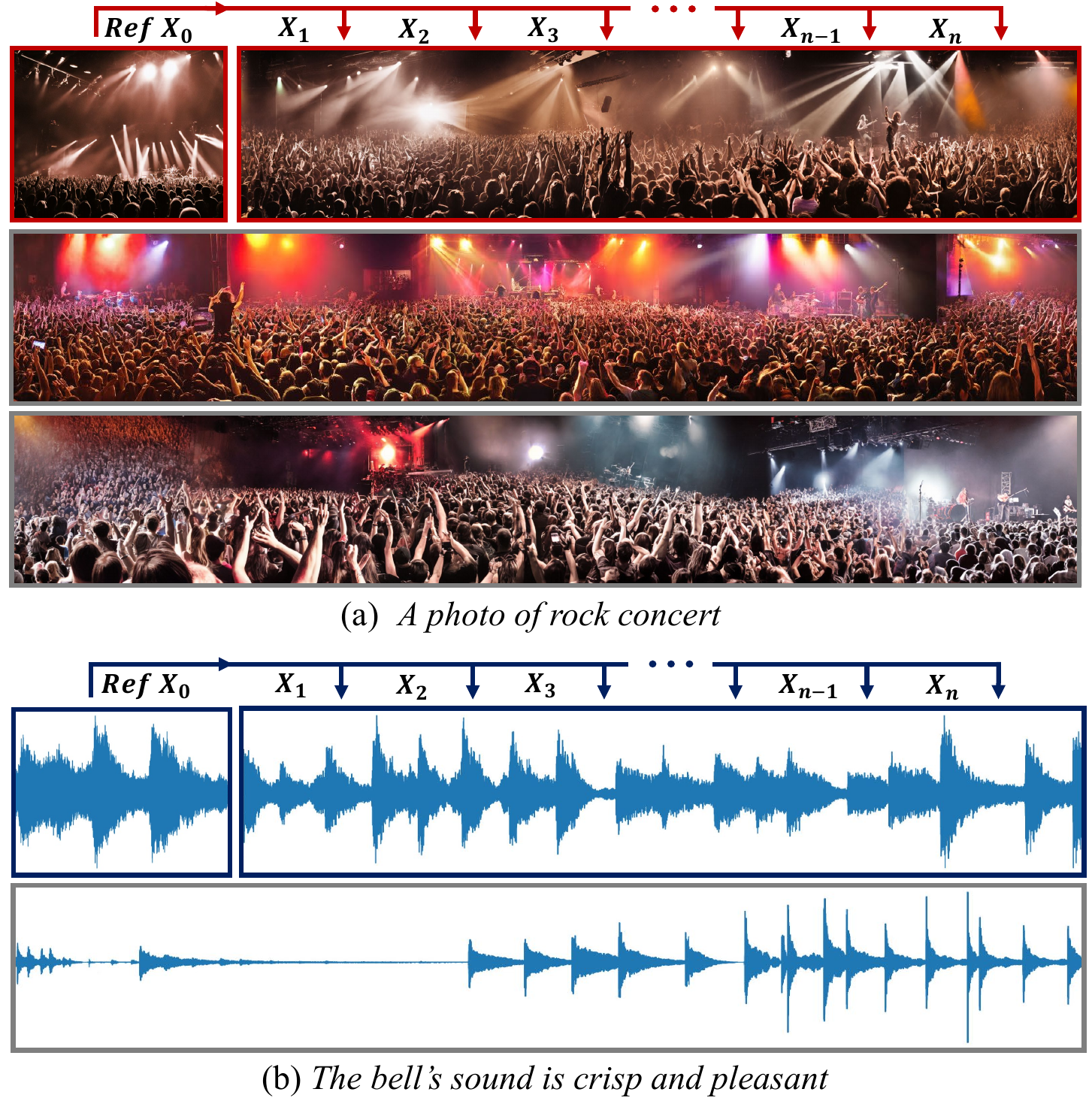}
\vspace{-15pt}
    \caption{Compared to unguided results in the last row of subfigures(a)(b), the panorama and waveform with Reference-Guided Swap (first row) demonstrate much more cross-view consistency, e.g., color and style for image, and timbre and SNR for audio. Considering the flattening order to the 1D sequence, in subfigure (a), we apply the swap along the row axis (first row) to achieve segment-level swaps and avoid cross-view repetition, compared to the pixel-wise swap along the column axis (middle row).}
\vspace{-15pt}
    \label{fig:small1}
\end{figure}

\begin{figure*}[!t]
    \centering
    \vspace{-20pt}
    \setlength{\belowcaptionskip}{0pt} 
    \includegraphics[width=2.1\columnwidth]{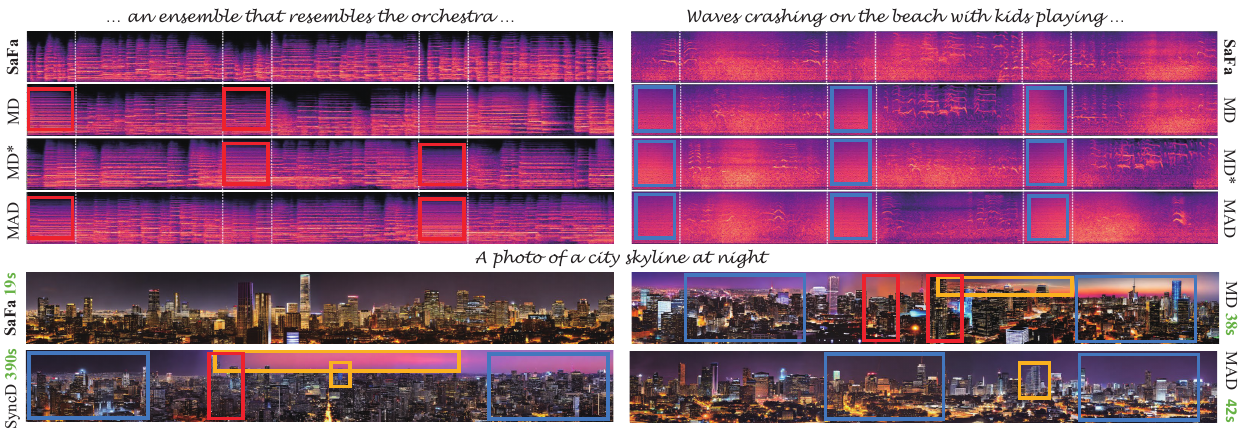}
    \vspace{-20pt}
    \caption{Qualitative comparisons. SaFa achieves better generation quality in both long audio generation and panoramas with \textcolor[RGB]{112,173,71}{high efficiency}. We highlight issues such as \textcolor[RGB]{255,0,0}{monotonous tails} and \textcolor[RGB]{46,117,182}{spectrum aliasing} in audio generation, \textcolor[RGB]{255,0,0}{ transition misalignment}, \textcolor[RGB]{46,117,182}{cross-view inconsistency } and \textcolor[RGB]{255,192,0}{ artifacts } in panorama generation caused by baseline methods.}
    \label{fig:qual}
    
\end{figure*}

As shown in Fig.~\ref{analysis} (c), we further compare the Fourier frequency analysis results of the overlap spectral latents produced by swap joint diffusion and standard joint diffusion. The results demonstrate that the latent swap operator effectively enhances the high-frequency components of the latent spectrum, closely aligning with the reference curve of the non-overlapping latent. As a result, the swap operation not only preserves spectral details but also significantly reduces aliasing artifacts in the overlapping regions. Moreover, the swap operation is also found to be well-suited for panoramic image generation, as illustrated in Fig.~\ref{fig:small1} and Fig.~\ref{fig:qual}.

\vspace{-10pt}
\paragraph{Reference-Guided Latent Swap}
\label{sec:Reference-Guided}


As mentioned in Introduction, previous work \cite{BarTal2023MultiDiffusionFD, lee2023syncdiffusion} relies on high overlap rate and test-time gradient optimization to mitigate cross-view inconsistency problem, bringing significant time cost and subview count. In contrast, we propose the unidirectional \textit{Reference-Guided Latent Swap} operation to efficiently achieve cross-view consistency in feed-forward manner. Specifically, in Fig.~\ref{fig:middle} (blues lines), for the early $r_{\text{guide}} \times T$ denoising steps, we guide the non-overlapping region of each subview trajectories \( M_i(J_t) \) by the same independent reference trajectory \( X_t^0 \) with a frame-level unidirectional swap operation to achieve cross-view consistency with swap operator $W_\text{refer}$ as:
\begin{equation}
\small
M_{i}(J_t) = W_\text{refer} \odot \text{Mid}(X_t^0) + (1-W_\text{refer}) \odot \text{ Mid }(X_t^{i})
\label{guide}
\end{equation}
where $\text{Mid}(\cdot)$ represents the region slice operation on the subview to obtain the non-overlapping region. We claim that the Reference-Guided Latent Swap can be considered a frame-level blended diffusion \cite{avrahami2022blended} that aligns the non-overlapping trajectory $ M_i(J_t) $ with the reference trajectory $x_t^0$, while maintaining coherence with nearby overlapping trajectories $I_{i-1,i}(J_t)$ and $I_{i,i+1}(J_t)$.  For the latter $ (1 -  r_{\text{guide}}) \times T $ of the denoising steps, the non-overlapping trajectories start from the similar intermediate denoised latent maps to simultaneously achieve cross-view consistency and avoid repetition, as illustrated on the left side of~\ref{fig:small1}. By adjusting the swap timing (early $r_{\text{guide}} \times T$ stage) and the swap interval ($w$ in Eq.~\ref{swapo}), we can achieve a trade-off of similarity and diversity for global coherence. As shown in Appendix Fig.~\ref{fig:ap2}, we observe an increase in similarity but a decrease in diversity as $ r_{\text{guide}}$ increases, and give a theoretical analysis in Appendix Sec.\ref{Theoretical}. As a result, we implement SaFa with a $r_{\text{guide}}$ of 0.3 as a similarity-diversity balance. As for the swap interval, we follow the implementation in Eq.~\ref{swapo}, adopting a frame-wise column swap with $w$ of 1. For image generation, in Fig.~\ref{fig:middle} right, considering the flattening order in the 1D token sequence\footnote{Spectrograms are flattened first along the frequency axis and then along the time axis, while images follow the reverse order.}, We adopt a row swap to achieve segment-wise swapping (Fig.\ref{fig:small1} (a), line 1) instead of a pixel-level swap (Fig.\ref{fig:small1} (a), line 2) in 1D token sequences, preventing excessive similarity due to the strong correlation between neighboring tokens in the attention operation.

\begin{algorithm}[t]
\small
\caption{Latent Swap Joint Diffusion}
\setlength{\baselineskip}{12pt} 
\textbf{Input}: Reference model $\Phi$,  subview mapping $\{F_i\}^n_{i=0}$, \\ conditions $\{y_i\}^n_{i=0}$, guidance step rate \textbf{$r_{\mathrm{refer}}$}, \\ region slice operations $\text{Right}(\cdot)$, $\text{Mid}(\cdot)$, and $\text{Left}(\cdot)$.\\
$J_T \sim P_{\mathcal{J}} \quad \triangleright$ noise initialization \\
\For{$t \gets T$ \KwTo 1}{
         $X_t^i = \Phi\big(F_i(J_{t+1}), y_i\big), \forall i \in[0,n]$\; 
          // Self-Loop Latent Swap $\triangleright$ Eq.~\ref{swap_equation} 
        $I_{i, i+1}(J_t) = \text{Swap}\big(\text{Left}(X^{i+1}_t), \text{Right}(X_t^i)\big), \forall i \in[n]$\; 
    \If{$t \geq (1-r_{\mathrm{guide}})\ \times T$}{
      //  Reference-Guided Latent Swap $\triangleright$ Eq.~\ref{guide} 
            $M_{i}(J_t) = \text{Swap}\big(\text{Mid}(X^0_t), \text{Mid}(X_t^i)\big), \forall i \in[n]$\; } 
    }
    
$\textbf{Output}: J_0 $ \\

// the implement of frame-level latent swap \\
\SetKwFunction{FSwap}{Swap} 
\SetKwProg{Fn}{Function}{:}{}
\Fn{\FSwap{$X_1, X_2$} }{ 
    $ W_\text{swap} = \mathbf{1}_n \otimes v_m,
v_m^{(i)} = \frac{1}{2} \left[ 1 - (-1)^{\left\lfloor \frac{i-1}{w} \right\rfloor} \right] $ $\triangleright$ Eq.~\ref{swapo}\\
    $X_{\text{new}} = W_\text{swap} \odot X_1 \ + (1-W_\text{swap}) \odot X_2$ \\ 
    $\textbf{Return}: X_{\text{new}}$\;
}
\label{alg:1}
\end{algorithm}

\vspace{-5pt}
\section{Experiment}
\label{sec:exp}

\subsection{Long-Form Audio Generation}
\label{sec:exp_audio}
\paragraph{Baselines}

We compare our approach with other joint diffusion methods in Tab.~\ref{tab:audio1}, including MultiDiffusion (MD) \cite{BarTal2023MultiDiffusionFD}, its enhanced version (MD*) \cite{polyak2024movie} with the triangular window, and Merge-Attend-Diffuse (MAD) \cite{Quattrini2024MergingAS}. SyncDiffusion \cite{lee2023syncdiffusion} is not implemented in audio generation, as LPIPS loss is observed insensitive to \textit{intermediate} denoised mel-spectra despite being sensitive to \textit{completely} denoised ones.


\begin{table*}[!t]
\centering
\footnotesize
\setlength{\tabcolsep}{2.3mm}

\vspace{-5pt}
\begin{tabular}{l|cccccc|cccccc}
\toprule[1pt]
\multirow{2}{*}{\textbf{Method}} & \multicolumn{6}{c|}{\textbf{DiT}}                                                                       & \multicolumn{6}{c}{\textbf{U-Net}}                                                                      \\
                                 & \textbf{FD↓}  & \textbf{FAD↓} & \textbf{KL↓}  & \textbf{CLAP↑} & \textbf{I-LPIPS ↓} & \textbf{I-CLAP↑} & \textbf{FD↓}  & \textbf{FAD↓} & \textbf{KL↓}  & \textbf{CLAP↑} & \textbf{I-LPIPS ↓} & \textbf{I-CLAP↑} \\ \hline
Reference                        & 2.92          & 0.22          & 0.74          & 0.54            & 0.39               & 0.86             & 4.12          & 0.27          & 1.15          & 0.53            & 0.43               & 0.79             \\
MAD   \cite{Quattrini2024MergingAS}                       & 12.77         & 7.56          & 0.86          & 0.51            & \cellcolor{pink}{0.32}      & 0.93             & 16.33         & 8.12          & 1.25          & 0.49            & \cellcolor{violet!20}{0.36}      & 0.88             \\
MD \cite{BarTal2023MultiDiffusionFD}                           & 11.31         & 6.41          & 0.81          & 0.51            & 0.36               & 0.91             & 14.29         & 7.06          & 1.18          & 0.50            & 0.41               & 0.86             \\
MD* \cite{polyak2024movie}                         & 9.79          & 5.09          & 0.77          & 0.52            & 0.36               & 0.92             & 11.24         & 5.53          & 1.12          & 0.51            & 0.40               & 0.88             \\
\textbf{SaFa}                    & \cellcolor{pink}{6.84} & \cellcolor{violet!20}{4.91} & \cellcolor{pink}{0.73} & \cellcolor{pink}{0.54}   & \cellcolor{violet!20}{0.34}      & \cellcolor{pink}{0.95}    & \cellcolor{violet!20}{7.88} & \cellcolor{violet!20}{4.27} & \cellcolor{pink}{1.11} & \cellcolor{violet!20}{0.53}   & \cellcolor{pink}{0.36}      & \cellcolor{pink}{0.92}    \\
\textbf{SaFa*}                   & \cellcolor{violet!20}{6.98} & \cellcolor{pink}{4.89} & \cellcolor{violet!20}{0.73} & \cellcolor{violet!20}{0.54}   & 0.36               & \cellcolor{violet!20}{0.94}    & \cellcolor{pink}{7.58} & \cellcolor{pink}{4.14} & \cellcolor{violet!20}{1.12} & \cellcolor{pink}{0.54}   & 0.39               & \cellcolor{violet!20}{0.90}    \\ \bottomrule[1pt]
\end{tabular}
\vspace{-10pt}
\caption{Quantitative comparisons on audio generation (including soundscape, sound effect and music). In SaFa*, only Self-Loop Swap is applied. MD* \cite{polyak2024movie} uses a triangular window achieving gradual transition. Pink and purple blocks represent the best and second-best results.
}
\label{tab:audio1}
\vspace{-15pt}
\end{table*}

\vspace{-10pt}
\paragraph{Experiment Settings}
As shown in Tab. \ref{tab:audio1}, all methods are implemented on two pretrained text-to-audio (TTA) models based on the AudioLDM \cite{liu2023audioldm} framework. One adopts AudioLDM’s original U-Net architecture and the other is based on a masked DiT architecture \cite{gao2023masked}. Both models incorporate a FLAN-T5 text encoder \cite{chung2024scaling} and a pretrained 2D spectrum-based VAE model \cite{liu2023audioldm}, and are trained on the same dataset and pipeline with variable length of audio, music and speech clips during 0.32s to 10.24s following Make-an-audio2\cite{Huang2023MakeAnAudio2T}. In inference stage, we use a DDIM sampler \cite{song2020denoising} with 200 denoising steps and a classifier-free guidance scale of 3.5. For vocoder, we employ HiFi-GAN \cite{kong2020hifi} to generate the audio samples from the mel-spectrogram. For quantitative experiments, we evaluate all the methods by generating 24s of audio at a 16kHz sample rate, combining three overlapping 10-second segments with an overlap rate $r_{\text{overlap}}$ of 0.2. For qualitative experiments, the duration of the segments is set to 8s to facilitate user studies. Nine text prompts are used (three soundscape, three sound effect and three music), as listed in the first nine audio prompts in Appendix.

\vspace{-10pt}
\paragraph{Evaluation Metrics}
Following AudioLDM \cite{liu2023audioldm}, Frechet Distance (FD) and Frechet Audio Distance (FAD) are used for quality and fidelity estimation (similar to FID score in image generation). KL divergence (KL) is also used at a pair level. To align with previous work \cite{BarTal2023MultiDiffusionFD,kim2024leveraging,lee2023syncdiffusion}, we first utilize the reference model to generate 500 10-seconds audio clips per prompt, obtaining the reference set. Since the objective metrics models are trained on 10-second audio clips, we sequentially extract 10-second segments from each long-form generated audio with a sliding window, obtaining multiple 500-sample evaluation subsets per prompt. Then we calculate FD, FAD, and KL scores between these subsets and the reference set, and average them to obtain the final score. As the reference, we also calculate these scores between two equal-sized random splits of the reference set. For semantic alignment, CLAP score is applied \cite{wu2023large}. Intra-LPIPS and Intra-CLAP (cosine similarity of audio CLAP embeddings) are used to estimate cross-view coherence by calculating internal similarity between 10s clips cropped from the slide window operation with overlap rate of 0.2. We claim that FD, FAD, KL, and Intra-CLAP scores are based on audio signals, while Intra-LPIPS is based on the Mel-spectrum.


\vspace{-10pt}
\paragraph{Quantitative Result}
As shown in Tab. \ref{tab:audio1}, in both DiT and U-Net models, SaFa consistently outperforms other methods significantly in semantic alignment (CLAP) and generation quality (FD, FAD, and KL). SaFa approaches reference-level performance in CLAP and KL, demonstrating the latent swap operator's superiority in preserving high-frequency details and avoiding spectrum aliasing compared to the averaging operator. Compared to SaFa* (with only Self-Loop Swap), SaFa achieves greater cross-view consistency (I-LPIPS and I-CLAP) through Reference-Guided Swap. MD* surpasses MD, consistent with \cite{polyak2024movie}, but remains inferior to SaFa. MAD employs block-wise averaging operations before each attention layer in the early stage, exacerbating spectrum aliasing and resulting in low quality. Additionally, it introduces a problem of position embedding repetition, resulting in monotonous and repetitive subviews with low I-LPIPS score. As shown in Tab.\ref{tab:audio2}, we further evaluate SaFa's stable performance on longer audio generation.

\vspace{-10pt}
\paragraph{Qualitative Result and User Study}
In Fig.~\ref{fig:qual} (top), SaFa preserves more high-frequency details without spectrum aliasing in the overlap compared with other methods. More qualitative comparisons are provided in Appendix Sec. \ref{sec:qual}. Furthermore, we conduct user studies on the generated audio samples to enhance evaluation reliability, collecting a total of 34 valid responses, where participants ranked the generated audio based on auditory quality and global consistency (including transition smoothness and cross-view consistency). The result aligns well with quantitative performance, showing a significant advantage for SaFa. Detail settings and results are available in Appendix Sec. \ref{user} and Fig.~\ref{fig:user1}.


\begin{table*}[!t]
\footnotesize
\centering
\vspace{-25pt}
\setlength{\tabcolsep}{2.35mm}
\begin{tabular}{l|cccccc|cccccc}
\toprule[1pt]
\multirow{2}{*}{\textbf{Method}} & \multicolumn{6}{c|}{\textbf{DiT}}                                                                     & \multicolumn{6}{c}{\textbf{U-Net}}                                                                    \\
                                 & \textbf{FD↓} & \textbf{FAD↓} & \textbf{KL↓} & \textbf{CLAP↑}  & \textbf{I-LPIPS ↓} & \textbf{I-CLAP↑} & \textbf{FD↓} & \textbf{FAD↓} & \textbf{KL↓} & \textbf{CLAP↑}  & \textbf{I-LPIPS ↓} & \textbf{I-CLAP↑} \\ \hline
SaFa (24s)                        & 6.84         & 4.91          & 0.73  & 0.54                    & 0.34               & 0.95             & 7.88         & 4.27          & 1.11   & 0.53                   & 0.36               & 0.92             \\
SaFa (48s)                        & 6.94         & 4.97          & 0.73  & 0.54                     & 0.35               & 0.94             & 7.61         & 4.10          & 1.08   & 0.54                   & 0.37               & 0.88             \\
SaFa (72s)                        & 6.98         & 4.99          & 0.72  & 0.54                    & 0.35               & 0.93             & 7.68         & 4.21           & 1.13 & 0.53                    & 0.37               & 0.89             \\ \bottomrule[1pt]
\end{tabular}
\vspace{-10pt}
\caption{SaFa maintains stable performance in longer audio generation at 24s, 48s, and 72s with both DiT and U-Net architectures.}
\vspace{-10pt}
\label{tab:audio2}
\end{table*}

\begin{table*}[!t]
\centering
\footnotesize
\setlength{\tabcolsep}{1.9mm}
\begin{tabular}{l|cccccc|cccccc}
\toprule[1pt]
\multirow{2}{*}{\textbf{Method}} & \multicolumn{6}{c|}{\textbf{DiT (SD 3.5)}}                                                                       & \multicolumn{6}{c}{\textbf{U-Net (SD 2.0)}}                                                                       \\
                                 & \textbf{FID↓} & \textbf{KID↓} & \textbf{CLIP↑} & \textbf{I-StyleL↓} & \textbf{I-LPIPS↓} & \textbf{Runtime↓} & \textbf{FID↓} & \textbf{KID↓} & \textbf{CLIP↑} & \textbf{I-StyleL↓} & \textbf{I-LPIPS↓} & \textbf{Runtime↓} \\ \hline
Reference                        & 28.19          & 0.01           & 32.57           & 5.49                & 0.60               & -                 & 33.37          & 0.01           & 31.60            & 8.72                & 0.73               & -                 \\
MD \cite{BarTal2023MultiDiffusionFD}                             & 24.50          & 8.12           & 32.37           & 2.58                & 0.59               & 103.85
            & \cellcolor{violet!20}{32.99} & \cellcolor{violet!20}{8.08}  & 31.76            & 3.08                & 0.69               & 37.71           \\
SyncD \cite{lee2023syncdiffusion}                           & 24.25          & 8.07           & 32.36           & 2.54                & 0.57               &  623.59
           & 44.58          & 19.98          & \cellcolor{violet!20}{31.84}   & \cellcolor{pink}{1.42}       & \cellcolor{pink}{0.55}      & 390.63        \\
MAD \cite{Quattrini2024MergingAS}                         & 65.10          & 55.73          & 31.79           & \cellcolor{pink}{0.67}       & \cellcolor{pink}{0.47}      & 85.25
           & 48.25          & 28.14          & 32.11            & 1.94                & \cellcolor{violet!20}{0.59}               & 41.82            \\
\textbf{SaFa}                    & \cellcolor{violet!20}{22.54} & \cellcolor{violet!20}{4.53}  & \cellcolor{pink}{32.45}  & \cellcolor{violet!20}{1.36}       & \cellcolor{violet!20}{0.56}      & \cellcolor{violet!20}{49.54}   & 34.71          & 9.91           & \cellcolor{pink}{31.84}   & \cellcolor{violet!20}{1.74}       & 0.61               & \cellcolor{violet!20}{19.15}    \\
\textbf{SaFa*}                   & \cellcolor{pink}{22.12} & \cellcolor{pink}{4.27}  & \cellcolor{violet!20}{32.39}  & 2.96                & 0.59               & \cellcolor{pink}{49.51}   & \cellcolor{pink}{32.43} & \cellcolor{pink}{6.97}  & 31.74            & 2.66                & 0.65               & \cellcolor{pink}{19.08}    \\ \bottomrule[1pt]
\end{tabular}
\vspace{-10pt}
\caption{Quantitative comparison on panorama generation with the resolution of 512 $\times$ 3200. In SaFa*, only Self-Loop Latent Swap is applied. KID and I-StyleL values are scaled by $10^3$. Pink and purple blocks represent the best and second-best results, respectively.}
\vspace{-15pt}
\label{tab:qual}
\end{table*}

\vspace{-10pt}
\paragraph{Comparison with Training-Based Methods}
We compare SaFa with SOTA training-based audio generation models includes AudioGen \cite{kreuk2022audiogen}, Stable Diffusion Audio  \cite{Evans2024FastTL} and Make-An-Audio2 \cite{Huang2023MakeAnAudio2T} on large-scale audio generation benchmark in Appendix Tab.~\ref{tab:ap1}. As a result, SaFa outperforms other methods across various length generation. Detailed settings and results are shown in Appendix Sec. \ref{training-based}.
\vspace{-10pt}
\subsection{Panorama Generation}
\paragraph{Experiment Settings}
We compare SaFa with MD \cite{BarTal2023MultiDiffusionFD}, MAD \cite{Quattrini2024MergingAS}, and SyncD \cite{lee2023syncdiffusion} on SD 2.0 (U-Net) \cite{rombach2022high} and SD v3.5 (MMDiT). Using the same six prompts from prior work \cite{BarTal2023MultiDiffusionFD, Quattrini2024MergingAS, lee2023syncdiffusion} in Appendix Fig.~\ref{fig:image_0} to ~\ref{fig:image_5}, we generate 500 panorama images per prompt (resolution: 512 × 3200, subviews: 512 × 640). Following previous work \cite{BarTal2023MultiDiffusionFD, lee2023syncdiffusion, Quattrini2024MergingAS}, we implement other methods with optimized settings at a $r_{\text{overlap}}$ of 0.8. With explicit guidance in the non-overlap region, we implement a much lower $r_{\text{overlap}}$ of 0.2 to achieve higher efficiency with fewer subviews and maintain high quality. For SaFa, with explicit guidance in the non-overlap region, we use a much lower $r_{\text{overlap}}$ of 0.2 to improve efficiency while maintaining high quality, requiring fewer subviews.

\vspace{-10pt}
\paragraph{Evaluation Metrics}
FID \cite{heusel2017gans} and KID \cite{binkowski2018demystifying} measure fidelity and diversity. For each reference model (SD 2.0 or SD 3.5), we generate five hundred $512 \times 512$ subview images per prompt as the reference set, and the same number for panorama with each joint diffusion methods. Correspondingly sized subviews are sequentially cropped from five equally divided regions of the panorama to construct the evaluation datasets. FID and KID scores are computed between the evaluation and reference sets, with the reference scores calculated between two equal random splits of the reference set. I-LPIPS \cite{zhang2018unreasonable} and Intra-StyleL (I-StyleL) \cite{gatys2016image} assess cross-view consistency using 10 subview pairs cropped from each panorama. As a reference, 1,000 pairs are randomly selected from the reference dataset to compute the average I-LPIPS and I-StyleL scores. The Mean CLIP score (mCLIP) \cite{hessel2021clipscore} evaluates semantic alignment, averaged over six prompts. Runtime measures the total time required to generate a complete panorama. All experiments are conducted using the same A100 GPU and PyTorch version.

\vspace{-10pt}
\paragraph{Quantitative Result}
In Tab.~\ref{tab:qual}, compared to MD with the averaging operator, SaFa* demonstrates better generation quality (FID, KID) and global coherence (I-StyleL, I-LPIPS), highlighting the effective adaptation of the latent swap operator to image generation. With Reference-Guided Swap, SaFa further improves cross-view consistency (I-StyleL, I-LPIPS) with a negligible increase in time. Additionally, SaFa achieves significantly lower FID and KID scores and is $2 \sim 20 \times $ faster than MAD and SynD, demonstrating its high quality and efficiency. We note a gap in I-LPIPS when compared to SyncD, as it is directly optimized on LPIPS loss. MAD shows less model generalization capability, performing much worse in DiT (full of transformer layers) than U-Net due to its reliance on the self-attention layer’s capability for long sequences. Moreover, it causes repetition problem with much higher I-StyleL and I-LPIPS in DiT that is sensitive to position encoding. 


\vspace{-10pt}
\paragraph{Qualitative Result and User Study}
As shown in Fig.~\ref{fig:qual}, SaFa achieves better generation quality with significantly reduced time cost compared to MD and SynD, which still exhibit transition misalignment and lack cross-view consistency. Compared to MAD, SaFa achieves comparable performance with higher time efficiency and model generalizability. More qualitative results are provided in Appendix Sec. \ref{user}. Furthermore, user studies confirm the quantitative results in generation quality and cross-view coherence, as shown in Appendix Fig.~\ref{fig:user3}.

\vspace{-5pt}
\section{Conclusion}
\vspace{-5pt}
In this paper, we present SaFa, a simple yet efficient latent swap framework that employs two fundamental swap operators to generate seamless and coherent long-form audio. Compared to previous techniques, SaFa is more adaptable to various modality tasks (long audio and even panorama images) across different diffusion architectures. As a high-performance alternative to the averaging operation, this operator can be widely applied in existing joint diffusion methods to achieve state-of-the-art performance without incurring additional time or computational cost. For future research, the practicality of SaFa for 1D wave-based VAE latents or other discrete token-based representations requires further investigation, though it is limited by the current T2A models.

\clearpage




\bigskip
{
    \small
    \bibliographystyle{ieeenat_fullname}
    \bibliography{main}
    
}
\clearpage
\section {More Quantitive Experiments}
\subsection{Comparison with Training-Based Methods}
\label{training-based}
We further compare our method with more training-based long audio generation models, including both diffusion models and language models. Although strictly speaking, the absolute performance between models of varying sizes and trained on different datasets seems incomparable, the relative performance degradation of each model with increasing audio generation length can highlight the strengths and weaknesses of these methods for long-generation tasks.
\vspace{-10pt}
\paragraph{Baselines}
The training-based baselines include: (1) \textit{AudioGen} \cite{kreuk2022audiogen} : An autoregressive model based on learned discrete audio representations, inherently supporting ultra-long audio generation. (2) \textit{Stable Diffusion Audio} (SD-audio) \cite{Evans2024FastTL}: A diffusion model trained on a fixed 96-second window size with long audio, generating variable-length outputs through end-cutting. Although the open-source version is trained on a 47-second window, longer audio can still be generated by customizing the initial noise size. (3) \textit{Make-An-Audio2} (Make2) \cite{Huang2023MakeAnAudio2T}: A diffusion model trained on variable-length window sizes, with audio lengths ranging from 0 to 20 seconds. It supports a maximum length of 27 seconds, constrained by the learnable positional encoding limit. For SaFa, we implement it on Make-An-Audio2 for a clearer comparison, following the settings described in Section \ref{sec:exp_audio}.


\vspace{-10pt}
\paragraph{Evaluation Settings}
We evaluate these four methods using a large-scale benchmark, AudioCaps \cite{kim2019audiocaps}, whose test set includes 880 ground-truth samples collected from YouTube videos. The target generation lengths are set to 32, 64, and 96 seconds. For SaFa, these outputs are formed by concatenating 4, 8, and 12 audio clips of 10 seconds, respectively. As in Section \ref{sec:exp_audio}, we use FD, FAD, KL, and mCLAP to assess the generation quality and semantic alignment of the generated audio. Following previous work \cite{Evans2024FastTL}], we apply a 10-second sliding window operation with an 8-second step on long audio samples and further evaluate them with AudioCaps test set.

\begin{table}[!t]
\centering
\setlength{\tabcolsep}{2.0mm}
\begin{tabular}{lcccc}
\toprule[1pt]
\textbf{Method}              & \textbf{FD↓}   & \textbf{FAD↓} & \textbf{KL↓}  & \textbf{mCLAP↑} \\ \hline
SD-audio (10s)             & 38.23 & 6.20  & 2.19 & 0.40   \\
SD-audio (32s)             & 25.52 & 6.43 & 2.24 & 0.37   \\
SD-audio (64s)             & 25.82 & 6.12 & 2.25 & 0.35   \\
SD-audio (96s)             & 30.11 & 6.54 & 2.38 & 0.33   \\ \hline
AudioGen (10s)      & 16.88 & 4.36 & 1.52 & 0.55   \\
AudioGen (32s)      & 18.54 & 4.81 & 1.71 & 0.50   \\
AudioGen (64s)      & 19.53 & 5.02 & 1.76 & 0.50   \\
AudioGen (96s)      & 18.88 & 5.44 & 1.78 & 0.49   \\ \hline
Make2 (10s)         & 14.37 & 1.12 & 1.28 & 0.57   \\
Make2 (27s)         & 18.49 & 2.26& 1.55  & 0.49   \\ \hline
\textbf{SaFa (32s)} & 15.21 & 1.45 & 1.25 & 0.57   \\
\textbf{SaFa (64s)} & 15.14 & 1.25 & 1.24 & 0.57   \\
\textbf{SaFa (96s)} & 15.36 & 1.33 & 1.25 & 0.57   \\
\bottomrule[1pt]
\end{tabular}
\vspace{-5pt}
\caption{Quantitative Comparison with Training-Based Variable-Length Audio Generation Models.}
\vspace{-15pt}
\label{tab:ap1}
\end{table}

\vspace{-10pt}
\paragraph{Results}
As shown in Table \ref{tab:ap1}, Make2, the SOTA diffusion-based audio generation model, demonstrates excellent performance in 10-second audio generation. However, it shows significant performance degradation when generating its maximum-length output of 27 seconds, as most training audio clips are under 20 seconds, and it lacks adaptation to longer unseen lengths. In contrast, our SaFa (32s) method maintains high performance in terms of KL and mCLAP, with only minor degradation observed in FD and FAD compared with the reference model Make2 (10s). Moreover, SaFa consistently delivers strong performance for 32-, 64-, and 96-second generation tasks with minimal degradation. As for AudioGen, the SOTA LM audio generation model, its architecture is inherently suited for generating longer audio compared to diffusion models, its performance degrades significantly as the generation length increases from 10 seconds to 96 seconds, accompanied by substantial increases in memory and time costs. For SD-audio, improved FD performance is observed when increasing the generation length from 10 to 32 seconds, likely due to the majority of its training data being focused on longer durations. However, other metrics consistently decline from 10 to 96 seconds, although the degradation is less pronounced compared to AudioGen. This highlights the robustness of diffusion models in generating longer outputs within their maximum training window.

\begin{figure*}[t]
    \centering
    \includegraphics[width=1.8\columnwidth]{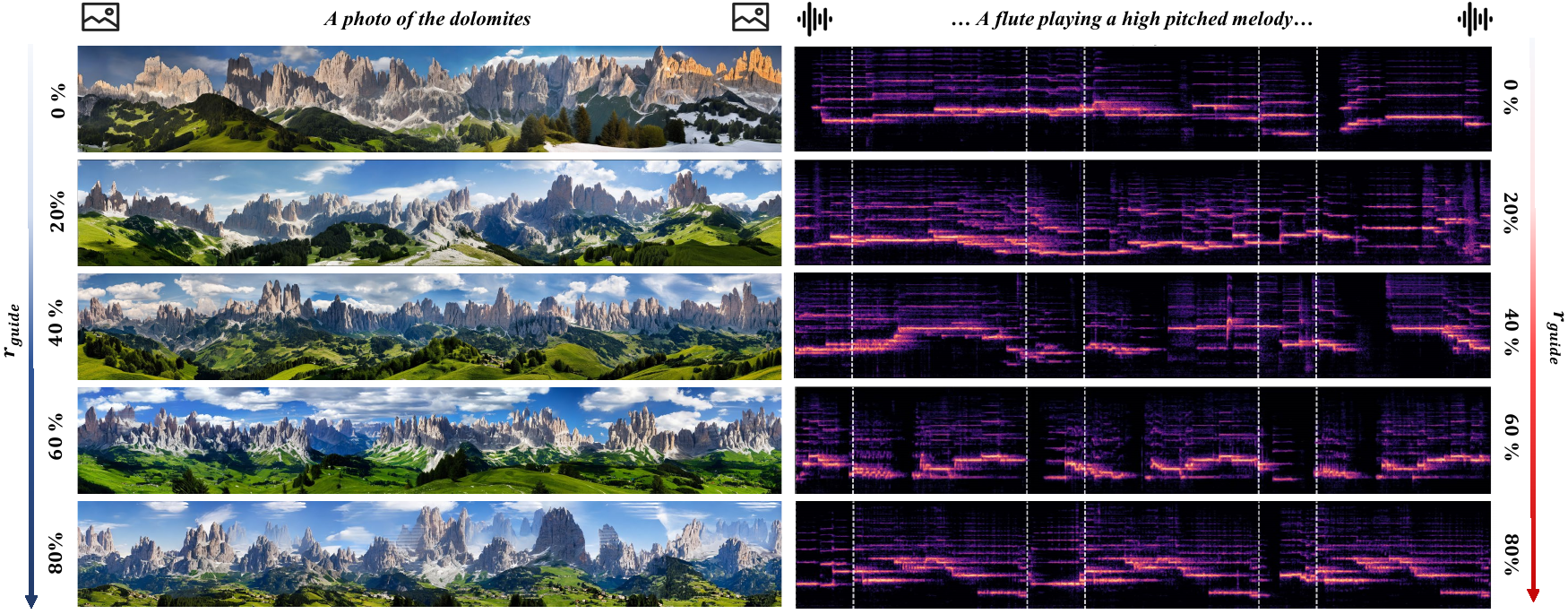}
    \caption{The effect of the trajectory guidance rate $r_\text{guide}$ in Reference-Guided Swap on long spectrum and panorama generation.}
    \vspace{-10pt}
\label{fig:ap2}
\end{figure*}

\subsection{Joint Diffusion on Open-Source Checkpoint}
In this subsection, we discuss several design flaws in existing open-source audio generation models that limit the application of training-free methods, such as the joint diffusion method. In this way, we show our audio generation model as a potential contribution to advancing training-free approaches in audio generation. 

\vspace{-5pt}
\paragraph{Adaptation on Existing T2A Models} Specifically, AudioLDM \cite{liu2023audioldm} and Tango \cite{ghosal2023tango} are trained with a fixed 10.24-second window, padding shorter clips with zeros or truncating longer clips. This flexible training pipeline causes unexpected end silence in generated audios. Consequently, implementing joint diffusion methods with these models often results in sudden silence in the overlap regions. Stable Diffusion Audio  \cite{Evans2024FastTL} is also trained with a fixed 96-second window and generates variable-length outputs by truncation, making it similarly challenging to adapt for joint diffusion methods. In comparison, Make-An-Audio2 follows a training pipeline similar to ours, using variable-length audio without excessive padding. It organizes samples into different buckets based on the length during training, randomly selecting samples from the same bucket within each batch. However, we observe some anomalous phenomena when applying Make2 with joint diffusion methods.

\begin{table}[!t]
\small
\centering
\vspace{5pt}
\renewcommand{\arraystretch}{0.95}
\setlength{\tabcolsep}{3.8mm}
\begin{tabular}{lcccc}
\toprule[1pt]
\textbf{Method} & \textbf{FD↓} & \textbf{FAD↓} & \textbf{KL↓} & \textbf{mCLAP↑} \\ \hline
Make2         & \underline{18.01}        & \underline{2.01}          & \underline{1.49}         & \underline{0.50}            \\
MD           & 65.28        & 17.70          & 3.22         & 0.24            \\
MAD          & 62.53        & 16.88         & 3.05         & 0.26            \\
\textbf{SaFa}   & \textbf{15.36}        & \textbf{1.32}          & \textbf{1.27}         & \textbf{0.57}            \\ 
\bottomrule[1pt]
\end{tabular}
\vspace{-5pt}
\caption{Quantitative comparisons of joint diffusion on 24-second audio generation on Make-An-Audio2 \cite{Huang2023MakeAnAudio2T}.}
\vspace{-10pt}
\label{tab:ap2}
\end{table}

\begin{table}[!t]
\small
\setlength{\tabcolsep}{3.8mm}
\setcounter{table}{7}
\renewcommand{\arraystretch}{0.95}
\centering
\begin{tabular}{lcccc}
\toprule[1pt]
\textbf{Method} & \textbf{FAD↓} & \textbf{FD↓}   & \textbf{KL↓}  & \textbf{mCLAP↑} \\ \hline
SD-audio        & \underline{6.44}          & \underline{25.26}    & \underline{2.18}          & \underline{0.37}            \\
MD              & 7.06          & 38.78          & 2.24          & 0.35            \\
MAD             & 7.56          & 38.9           & 2.21          & 0.35            \\
\textbf{SaFa}   & \textbf{4.19} & \textbf{24.40} & \textbf{1.96} & \textbf{0.41}       \\
\bottomrule[1pt]
\end{tabular}
\vspace{-5pt}
\caption{Quantitative comparisons of join diffusion on 24-second audio generation on Stable Diffusion Audio  \cite{Evans2024FastTL}.}
\vspace{-20pt}
\label{tab:stable}
\end{table}

\begin{figure}[!t]
    \centering
    \vspace{-5pt}
    \includegraphics[width=1.0\columnwidth]{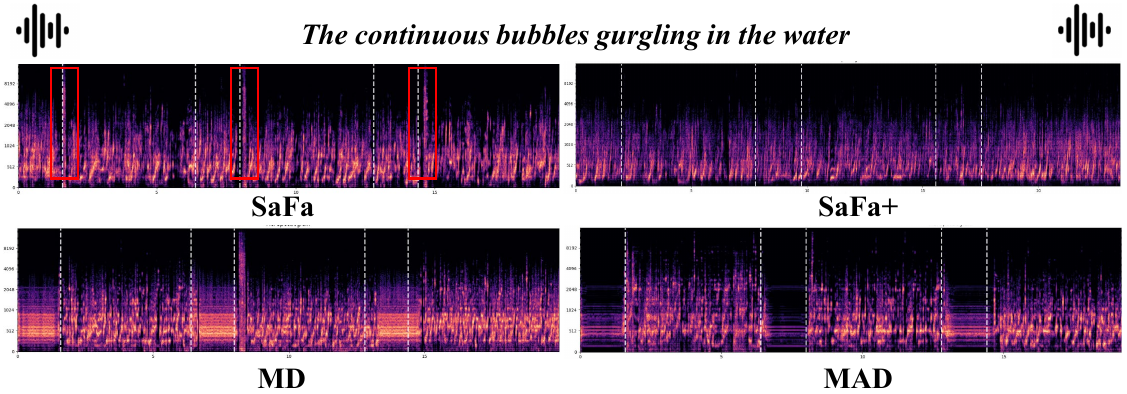}
    \vspace{-20pt}
    \caption{The long-form spectrum generated by various joint diffusion methods based on Make-An-Audio2. }
    \vspace{-15pt}
\label{fig:ap1}
\end{figure}

\vspace{-15pt}
\paragraph{Comparison on 1D Convolution VAE Latent} As shown in Figure \ref{fig:ap1}, when applying the joint diffusion method to Make2, short abrupt transitions appear at the end of each overlap region. Although SaFa significantly improves blending and generation quality compared to MD and MAD, these abrupt transitions still persist. Through experiments, we identify two main causes of this issue:
(1) The VAE latent map of Make-An-Audio2 is sensitive to the last token from an adjacent subview. To mitigate this, we apply Self-loop Swap with a five-token forward shift on the overlap regions. (2) Its VAE model is less robust to linear operations on the latent map compared to AudioLDM \cite{liu2023audioldm}. By performing concatenation at $t=0$ on the mel-spectrogram rather than on the latent map, we effectively resolve this issue. As a result, the improved method, SaFa+, performs well in Figure \ref{fig:ap1}.

For quantitative comparison in Table \ref{tab:ap2}, our method, SaFa, significantly outperforms Make2 and other joint diffusion methods across all metrics for 24-second generation tasks.

\paragraph{Comparison on Waveform VAE Latent} As shown in Table \ref{tab:stable}, we also compare SaFa with other joint diffusion methods on waveform VAE latents using the open-source checkpoint from SD-Audio. We note that although we restrict the initial latent maps on 10 seconds to adapt joint diffusion method (while the model was trained on fixed 96-second latent maps), this unoptimized setting does not compromise fairness. As a result, SaFa also significantly outperforms existing methods when evaluated on AudioCaps with SD-audio.

\begin{figure*}[t]
    \centering
    \includegraphics[width=1.6\columnwidth]{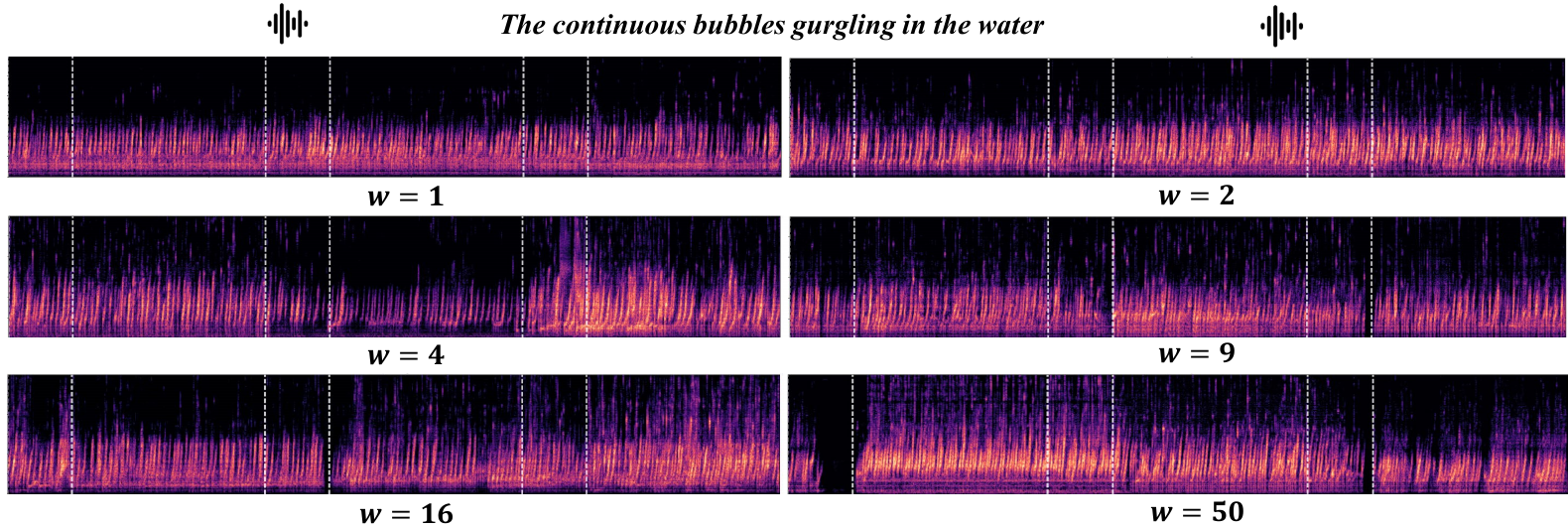}
    \caption{The effect of the swap interval $w$ (Eq. \ref{swapo} ) of Self-Loop Latent Swap on spectrum generation. Better transition is achieved with lower values of $w$, 1 or 2, which indicate a high swap frequency between two step-wise differential trajectories to enhance the high-frequency component in the denoised mel-spectrum with better-blender transitions.}
    \vspace{-15pt}
    \label{fig:ap3}
\end{figure*}

\begin{figure}[t]
    \centering
    \includegraphics[width=0.9\columnwidth]{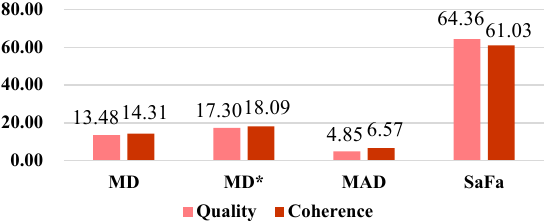}
    \vspace{-5pt}
    \caption{User study results on audio generation.}
      \vspace{-10pt}
\label{fig:user1}
\end{figure}

\subsection{Effect of Guidance Rate and Swap Interval}
In Figure\ref{fig:ap2}, we further demonstrate the progressive transition from cross-view diversity to similarity by varying $r_\text{guide}$ in both mel-spectrum and panorama generation using Reference-Guided Latent Swap. All other settings for SaFa remain consistent with Section \ref{sec:exp}. As shown in Figure \ref{fig:ap2}, using an appropriate trajectory guidance rate $r_\text{guide}$, 20\% to 40\%, results in unified cross-view coherence while preserving the diversity of local subviews. However, as the guidance rate $r_\text{guide}$ increases beyond 60\%, excessive repetition and artifacts begin to appear. This occurs because Reference-Guided Swap is a unidirectional operation, where the denoising process of the reference view is independent and unaffected by each subview. Consequently, it does not adapt as seamlessly to subviews in the later stages as the bidirectional Self-Loop Swap operation does. This is also one of the reasons why we restrict Reference-Guided Swap to the early denoising stages.


To further explore the effects of the swap interval $w$ (in Eq. \ref{swapo}), we apply the Self-Loop Latent Swap with various $w$ values in spectrum generation, as shown in Figure \ref{fig:ap3}. We observe that using a small swap interval (1 or 2), corresponding to higher swap frequencies, produces smoother transitions. Conversely, larger $w$ values indicate larger swap units, resulting in less seamless transitions between subviews. This outcome aligns with the high-frequency variability of mel tokens, leading us to default the Self-Loop Latent Swap to frame-level operations with $w=1$ for optimal performance.

\subsection{Length Adaptation on Panorama Generation}

In Table \ref{tab:length}, We utilize SD 2.0 model to estimate performance of SaFa on panorama images with resolutions of 512 $\times$ 1600, 512 $\times$ 3200, and 512 $\times$ 4800. As a result, SaFa maintains stable and great performance across all evaluated metrics in different length output.

\section{User Study}
\label{user}
For subjective evaluation, we randomly select samples from the qualitative results of the top four methods in audio and panorama generation for user studies. We use the same notation as in Section \ref{sec:exp_audio}. Specifically, SaFa is compared with MD, MD*, and MAD for audio generation, while for panorama generation, SaFa is compared with MD, MAD, and SyncD. For each task, we randomly select 30 parallel comparison groups (plus 2 additional pairs as a vigilance group) from the four compared methods, evenly distributed across six prompts. 
\begin{figure}[t]
    \centering
    \includegraphics[width=0.9\columnwidth]{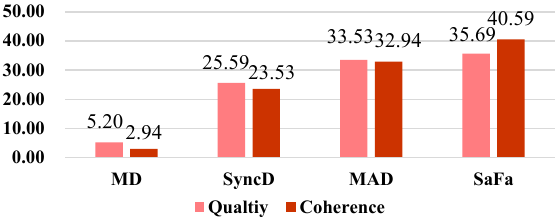}
    \caption{User study results on panorama generation.}
\label{fig:user3}
\end{figure}
We recruit 39 participants with basic machine learning knowledge but no prior familiarity with the research presented in this paper. Each participant is required to select the best sample from each of the 32 groups based on two evaluation dimensions: generation quality and global coherence. Semantic alignment is not considered, as most samples align well with the prompts semantically and cannot be easily distinguished in this regard. We ultimately collect 34 valid responses out of 39 participants. The results indicate that SaFa consistently outperforms the baseline methods, achieving superior human preference scores across both evaluation dimensions. Figure \ref{fig:user1} highlight the significant preference of human evaluators for SaFa in both quality and coherence assessments of audio generation. This preference stems from the swap operator's enhanced adaptability to the inherent characteristics of spectral data, which lacks the typical global structural features or contours present in images. Meanwhile in Figure \ref{fig:user3}, with significantly faster inference speeds and relying solely on fixed self-attention windows, SaFa achieves comparable performance to SyncD and MAD in the subjective evaluation of panorama generation.

\begin{table}[!t]
\centering
\small
\setlength{\tabcolsep}{1mm}
\renewcommand{\arraystretch}{0.95}
\begin{tabular}{lccccccc}
\toprule[1pt]
\textbf{Method}                  & \textbf{CLIP} $\uparrow$  & \textbf{FID} $\downarrow$ & \textbf{KID} $\downarrow$ & \textbf{I-LPIPS} $\downarrow$ & \textbf{I-StyleL} $\downarrow$ \\ \hline
SaFa (1600)          & 31.88                   & 34.47  & 9.71    & 0.59          & 1.67       \\
SaFa (3200)          & 31.84                  & 34.71    & 9.91   & 0.61         & 1.74        \\
SaFa (4800)          & 31.88                  & 34.97    & 10.68    & 0.62         & 1.78        \\
\bottomrule[1pt]
\end{tabular}

\vspace{-5pt}
\caption{Length adaptation of SaFa on panorama generation.}
\vspace{-15pt}
\label{tab:length}
\end{table}

\clearpage
\onecolumn 
\section{Theoretical Analysis of Refer-Guided Swap for Cross-View Similarity-Diversity Balance}
\label{Theoretical}
Reference-Guided Latent Swap improves cross-view consistency comparing with independent denoising process with reference model directly. When SD-2.0 \cite{rombach2022high} is employed as the reference model $\Phi$, we have the following proposition, which describes the difference between two updated samples from arbitrary starting points $\mathbf{x}_{t_2}^{(1)}, \mathbf{x}_{t_2}^{(2)}\in \mathcal{X}$: 

\paragraph{Proposition}
\label{Proposition}
Recall that the approximated reversed VP-SDE \cite{song2020score} used for conditionally generation in SD-2.0 is:

\begin{equation}
\textrm{d} \mathbf{x}=\left[-\frac{1}{2}\beta(t)\mathbf{x} - \beta( t )s_{\theta}(\mathbf{x},t,y)  \right] \textrm{d} t+\sqrt{\beta( t )} \textrm{d} \tilde{\mathbf{w}},
\end{equation}

where $s_{\theta}(\mathbf{x},t,y)$ is a estimation for $\nabla_{\mathbf{x}} \log p_{t} ( \mathbf{x} \vert y )$, and $\tilde{\mathbf{w}}$ is a Wiener process when time flows backwards from $t=1$ to $t=0$. Denote that $\Phi_{t_2\rightarrow t_1}(\cdot \vert y)$ is the the sampling procedure from $t_2$ to $t_1$ condition on $y$ in SD-2.0, and $\sigma^2_{t_2 \rightarrow t_1}=-\int_{t_2}^{t_1}\beta(u)\textrm{d}u$. Assume that $\forall \mathbf{x}\in \mathcal{X}, \forall t\in[0,1], \forall y\in \mathcal{Y},  \left\|s_{\theta}(\mathbf{x},t,y)\right\|_2\leq C$, then $\forall 0\leq t_1 < t_2 \leq 1$, $\forall \mathbf{x}_{t_2}^{(1)}, \mathbf{x}_{t_2}^{(2)}\in \mathcal{X}$, $\forall y\in \mathcal{Y}$, $\forall \delta\in(0,1)$, with probability at least $(1-\delta)$, 


\begin{equation}
\begin{aligned}
\label{eq1}
    \left\| \Phi_{t_2\rightarrow t_1}\left(\mathbf{x}_{t_2}^{(1)} \vert y\right) 
    -\Phi_{t_2\rightarrow t_1}\left(\mathbf{x}_{t_2}^{(2)} \vert y\right) \right\|_2^2 
    &\leq \exp(\sigma^2_{t_2 \rightarrow t_1}) 
    \left[\left\|\mathbf{x}_{t_2}^{(1)}-\mathbf{x}_{t_2}^{(2)}\right\|_2
    +2C \left\|\int_{t_2}^{t_1}\exp\left(-\frac{1}{2}\sigma^2_{t_2 \rightarrow s}\right)\beta(s)\,\textrm{d}s\right\|_2
    \right]^2 \\
    &\quad +2\sigma^2_{t_2 \rightarrow t_1} 
    \left( d+2\sqrt{d\cdot(-\log\delta)}+2\cdot(-\log\delta) \right).
\end{aligned}
\end{equation} where $d$ is the number of dimensions of $\mathbf{x}_{t_2}^{(1)}, \mathbf{x}_{t_2}^{(2)}, \mathbf{x}_{t_2}^{(ref)}$. 

\paragraph{Proof}
Using method of variation of parameters, solution for pre-mentioned SDE (1), $\Phi_{t_2\rightarrow t_1}(\mathbf{x}_{t_2} \vert y)$, can be written as

\begin{equation}
\Phi_{t_2\rightarrow t_1}(\mathbf{x}_{t_2} \vert y)\\
=\exp(\frac{1}{2}\sigma^2_{t_2 \rightarrow t_1})\left[\mathbf{x}_{t_2}
-\int_{t_2}^{t_1}\exp(-\frac{1}{2}\sigma^2_{t_2 \rightarrow s})\beta(s)s_{\theta}(\mathbf{x}_{s},s,y)\textrm{d}s\right]\\
+\int_{t_2}^{t_1}\sqrt{\beta( t )} \textrm{d} \tilde{\mathbf{w}},
\end{equation}

so for $\forall \mathbf{x}_{t_2}^{(1)}, \mathbf{x}_{t_2}^{(2)} \in \mathcal{X}$, we have: 

\begin{equation}
\begin{aligned}
& \left\| \Phi_{t_2\rightarrow t_1}\left(\mathbf{x}_{t_2}^{(1)} \vert y\right)-\Phi_{t_2\rightarrow t_1}\left(\mathbf{x}_{t_2}^{(2)} \vert y\right) \right\|_2^2 \\
= & \|\exp(\frac{1}{2}\sigma^2_{t_2 \rightarrow t_1})\Big[(\mathbf{x}_{t_2}^{(1)}-\mathbf{x}_{t_2}^{(2)})-\int_{t_2}^{t_1}\exp(-\frac{1}{2}\sigma^2_{t_2 \rightarrow s})\beta(s)(s_{\theta}(\mathbf{x}_{s}^{(1)},s,y)-s_{\theta}(\mathbf{x}_{s}^{(2)},s,y))\textrm{d}s_2^2\Big] \\
& +\int_{t_2}^{t_1}\sqrt{\beta( t )} \textrm{d} \tilde{\mathbf{w}}_1-\int_{t_2}^{t_1}\sqrt{\beta( t )} \textrm{d} \tilde{\mathbf{w}}_2\|_2^2 \\
= &\exp(\sigma^2_{t_2 \rightarrow t_1})\left\|\mathbf{x}_{t_2}^{(1)}-\mathbf{x}_{t_2}^{(2)}+\int_{t_2}^{t_1}\exp(-\frac{1}{2}\sigma^2_{t_2 \rightarrow s})\beta(s)(s_{\theta}(\mathbf{x}_{s}^{(1)},s,y)-s_{\theta}(\mathbf{x}_{s}^{(2)},s,y))\textrm{d}s\right\|_2^2  \\
&+\left\|\int_{t_2}^{t_1}\sqrt{\beta( t )} \textrm{d} \tilde{\mathbf{w}}_1-\int_{t_2}^{t_1}\sqrt{\beta( t )} \textrm{d} \tilde{\mathbf{w}}_2\right\|_2^2 \\
\leq &\exp(\sigma^2_{t_2 \rightarrow t_1})\left[  \left\|\mathbf{x}_{t_2}^{(1)}-\mathbf{x}_{t_2}^{(2)}\right\|_2+\left\|\int_{t_2}^{t_1}\exp(-\frac{1}{2}\sigma^2_{t_2 \rightarrow s})\beta(s)(s_{\theta}(\mathbf{x}_{s}^{(1)},s,y)-s_{\theta}(\mathbf{x}_{s}^{(2)},s,y))\textrm{d}s\right\|_2\right]^2  \\
& +2\left\|\int_{t_2}^{t_1}\sqrt{\beta( t )} \textrm{d} \tilde{\mathbf{w}}\right\|_2^2.
\end{aligned}
\end{equation}

From the assumption over $s_{\theta}(\mathbf{x},t,y)$, we have:
\begin{equation}
\begin{aligned}
    & \left\| \Phi_{t_2\rightarrow t_1}\left(\mathbf{x}_{t_2}^{(1)} \vert y\right) 
    -\Phi_{t_2\rightarrow t_1}\left(\mathbf{x}_{t_2}^{(2)} \vert y\right) \right\|_2^2 \\ 
    = & \Bigg\|\exp\left(\frac{1}{2}\sigma^2_{t_2 \rightarrow t_1}\right) 
    \Bigg[(\mathbf{x}_{t_2}^{(1)}-\mathbf{x}_{t_2}^{(2)}) 
    - \int_{t_2}^{t_1} \exp\left(-\frac{1}{2} \sigma^2_{t_2 \rightarrow s} \right) 
    \beta(s) \big( s_{\theta}(\mathbf{x}_{s}^{(1)},s,y) 
    - s_{\theta}(\mathbf{x}_{s}^{(2)},s,y) \big) \textrm{d}s \Bigg] \\
    & + \int_{t_2}^{t_1} \sqrt{\beta( t )} \, \textrm{d} \tilde{\mathbf{w}}_1 
    - \int_{t_2}^{t_1} \sqrt{\beta( t )} \, \textrm{d} \tilde{\mathbf{w}}_2 \Bigg\|_2^2 \\
    = & \exp\left(\sigma^2_{t_2 \rightarrow t_1}\right) 
    \left\| \mathbf{x}_{t_2}^{(1)}-\mathbf{x}_{t_2}^{(2)} 
    + \int_{t_2}^{t_1} \exp\left(-\frac{1}{2} \sigma^2_{t_2 \rightarrow s} \right) 
    \beta(s) \big( s_{\theta}(\mathbf{x}_{s}^{(1)},s,y) 
    - s_{\theta}(\mathbf{x}_{s}^{(2)},s,y) \big) \textrm{d}s \right\|_2^2  \\
    & + \left\| \int_{t_2}^{t_1} \sqrt{\beta( t )} \, \textrm{d} \tilde{\mathbf{w}}_1 
    - \int_{t_2}^{t_1} \sqrt{\beta( t )} \, \textrm{d} \tilde{\mathbf{w}}_2 \right\|_2^2 \\
    \leq & \exp\left(\sigma^2_{t_2 \rightarrow t_1}\right) 
    \left[  \left\|\mathbf{x}_{t_2}^{(1)}-\mathbf{x}_{t_2}^{(2)}\right\|_2 
    + \left\| \int_{t_2}^{t_1} \exp\left(-\frac{1}{2} \sigma^2_{t_2 \rightarrow s} \right) 
    \beta(s) \big( s_{\theta}(\mathbf{x}_{s}^{(1)},s,y) 
    - s_{\theta}(\mathbf{x}_{s}^{(2)},s,y) \big) \textrm{d}s \right\|_2 \right]^2  \\
    & + 2 \left\| \int_{t_2}^{t_1} \sqrt{\beta( t )} \, \textrm{d} \tilde{\mathbf{w}} \right\|_2^2.
\end{aligned}
\end{equation}

Then we complete the proof for Proposition \ref{Proposition}.

\paragraph{Corollary}
If we use introduce reference-guided latent swap operation before updating, by the definition of $\text{Swap}(\cdot)$, we have $\forall 0\leq t_1 < t_2 \leq 1$, $\forall \mathbf{x}_{t_2}^{(1)}, \mathbf{x}_{t_2}^{(2)}, \mathbf{x}_{t_2}^{(ref)} \in \mathcal{X}$, $\forall y\in \mathcal{Y}$, $\forall \delta\in(0,1)$, with probability at least $(1-\delta)$, 
\begin{equation}
\begin{aligned}
\label{eq2}
   & \left\| \Phi_{t_2\rightarrow t_1}\left(\text{Swap}(\mathbf{x}_{t_2}^{(ref)},\mathbf{x}_{t_2}^{(1)}) \vert y\right) 
    -\Phi_{t_2\rightarrow t_1}\left(\text{Swap}(\mathbf{x}_{t_2}^{(ref)},\mathbf{x}_{t_2}^{(2)}) \vert y\right) \right\|_2^2
\\& \leq   \exp\left(\sigma^2_{t_2 \rightarrow t_1}\right) 
    \Bigg[ \left\|(1-W_{\textrm{swap}})\odot(\mathbf{x}_{t_2}^{(1)}-\mathbf{x}_{t_2}^{(2)})\right\|_2   + 2C \left\| \int_{t_2}^{t_1} \exp\left(-\frac{1}{2} \sigma^2_{t_2 \rightarrow s}\right) 
    \beta(s) \,\textrm{d}s \right\|_2 \Bigg]^2  \\ 
    & + 2\sigma^2_{t_2 \rightarrow t_1} 
    \left( d + 2\sqrt{d \cdot (-\log\delta)} + 2 \cdot (-\log\delta) \right).
\end{aligned}
\end{equation}

Comparing with Eq.\ref{eq1}, Eq.\ref{eq2} have a tighter upper bound, since $\text{Swap}(\mathbf{x}_{t_2}^{(ref)},\mathbf{x}_{t_2}^{(1)}),\text{Swap}(\mathbf{x}_{t_2}^{(ref)},\mathbf{x}_{t_2}^{(2)})$ shares the same part $W_{\textrm{swap}}\odot\mathbf{x}_{t_2}^{(ref)}$. This indicates that within a fixed time interval $[t_1, t_2]$, performing a reference-guided swap operation on the initial points before updating the sample points helps improve the similarity of the results. 

According to Eq.\ref{eq1} and Eq.\ref{eq2}, we can trade-off between similarity and diversity by tuning $r_\text{guide}$. As $r_\text{guide}$ increases, the swap operation is employed more frequently applied during the sampling process, leading to higher similarity across subviews. Conversely, an increase in the $L_2$ distance between the final subview images signifies enhanced sample diversity.

\section{Further Qualitative Comparison}
\label{sec:qual}
\begin{figure*}[t]

    \centering
    \setlength{\belowcaptionskip}{-10pt} 
    \includegraphics[width=2.0\columnwidth]{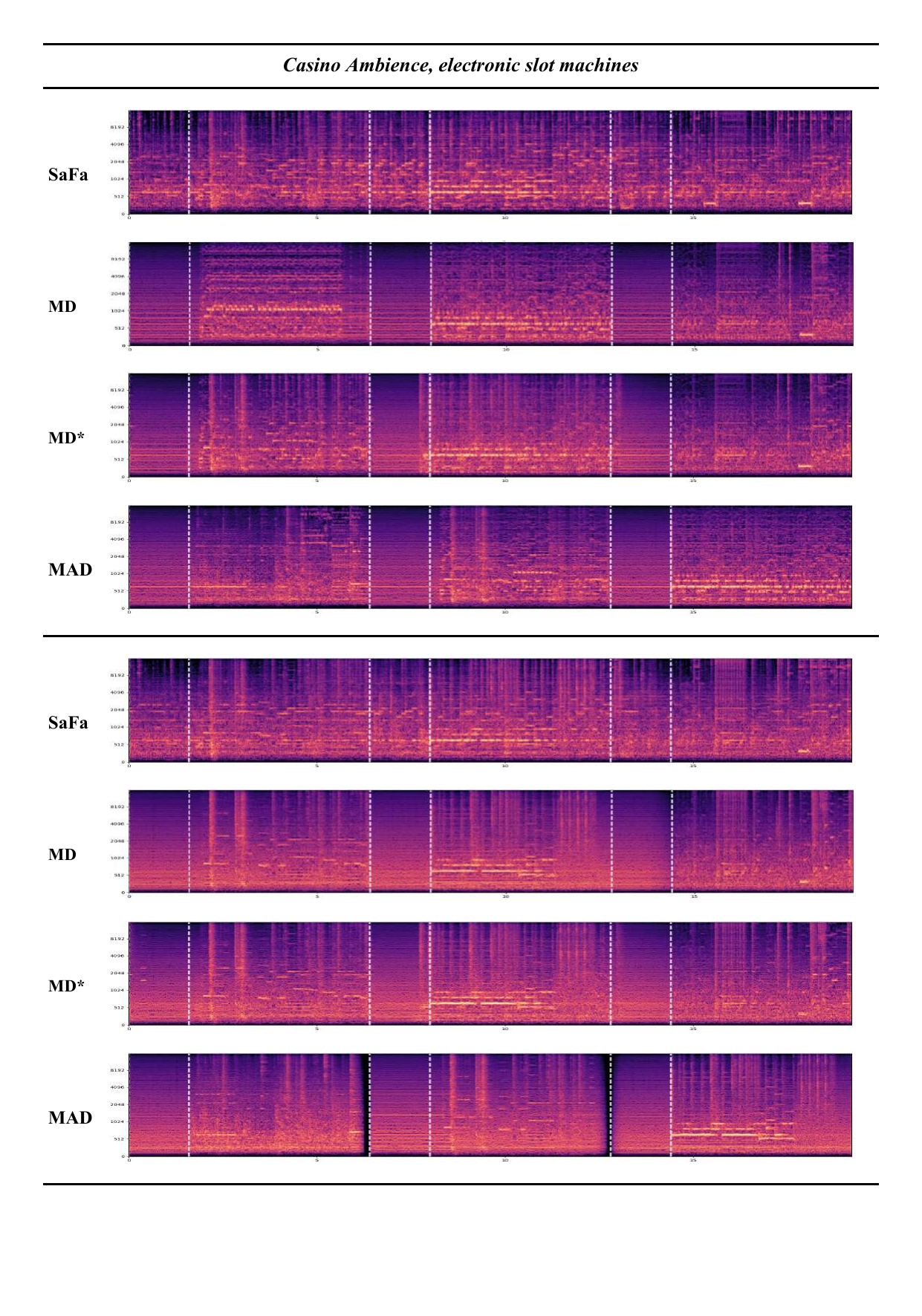}

\vspace{-60pt}
\caption{Qualitative comparison on soundscape generation. MD* represent an enhanced MD method with triangular windows.}
\label{fig:audio_0}
\end{figure*}

\begin{figure*}[t]
    \centering
    \setlength{\belowcaptionskip}{-10pt} 
    \includegraphics[width=2.0\columnwidth]{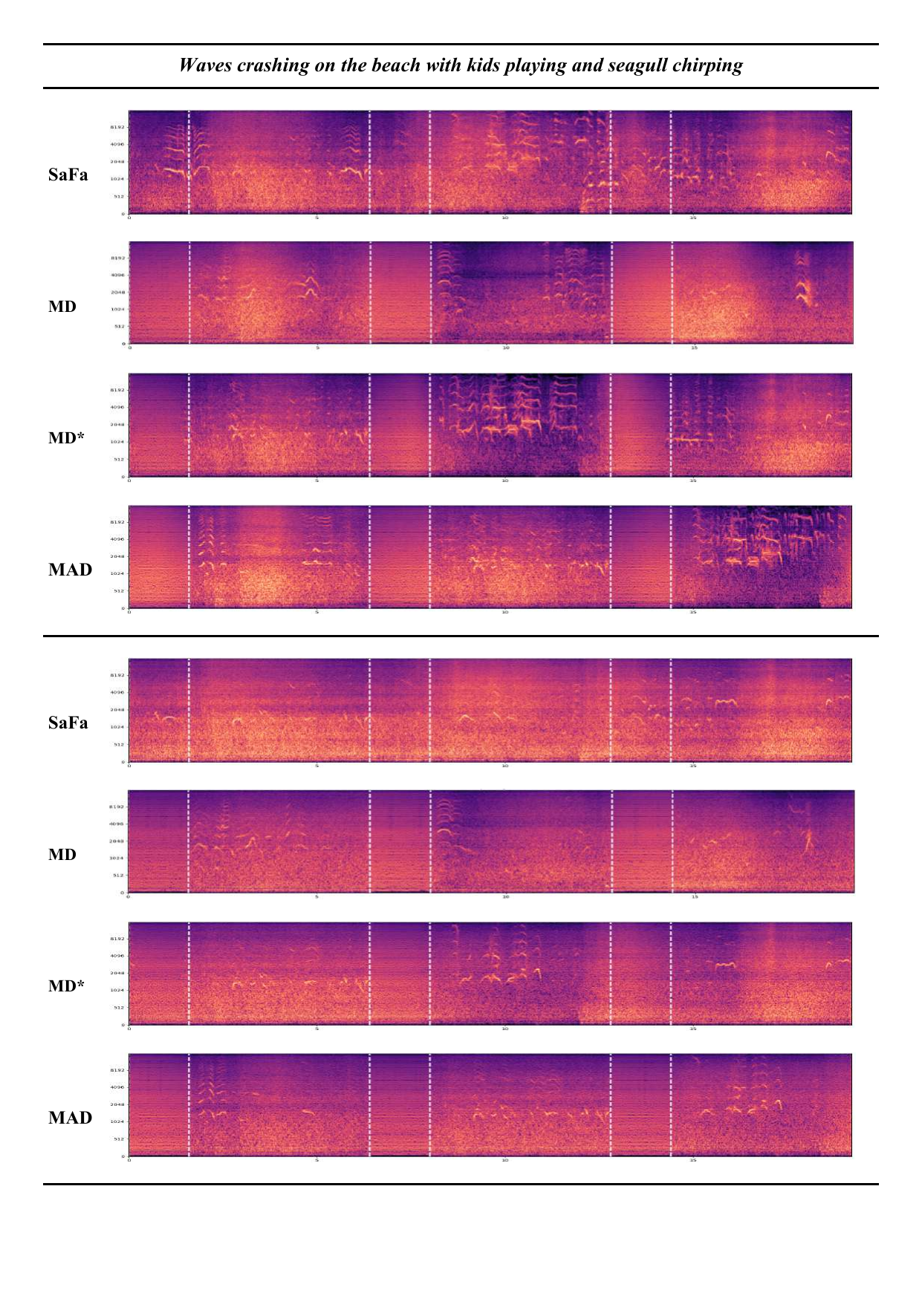}

\vspace{-60pt}
\caption{Qualitative comparison on soundscape generation. MD* represent an enhanced MD method with triangular windows.}
\label{fig:audio_5}
\end{figure*}

\begin{figure*}[t]
    \centering
    \setlength{\belowcaptionskip}{-10pt} 
    \includegraphics[width=2.0\columnwidth]{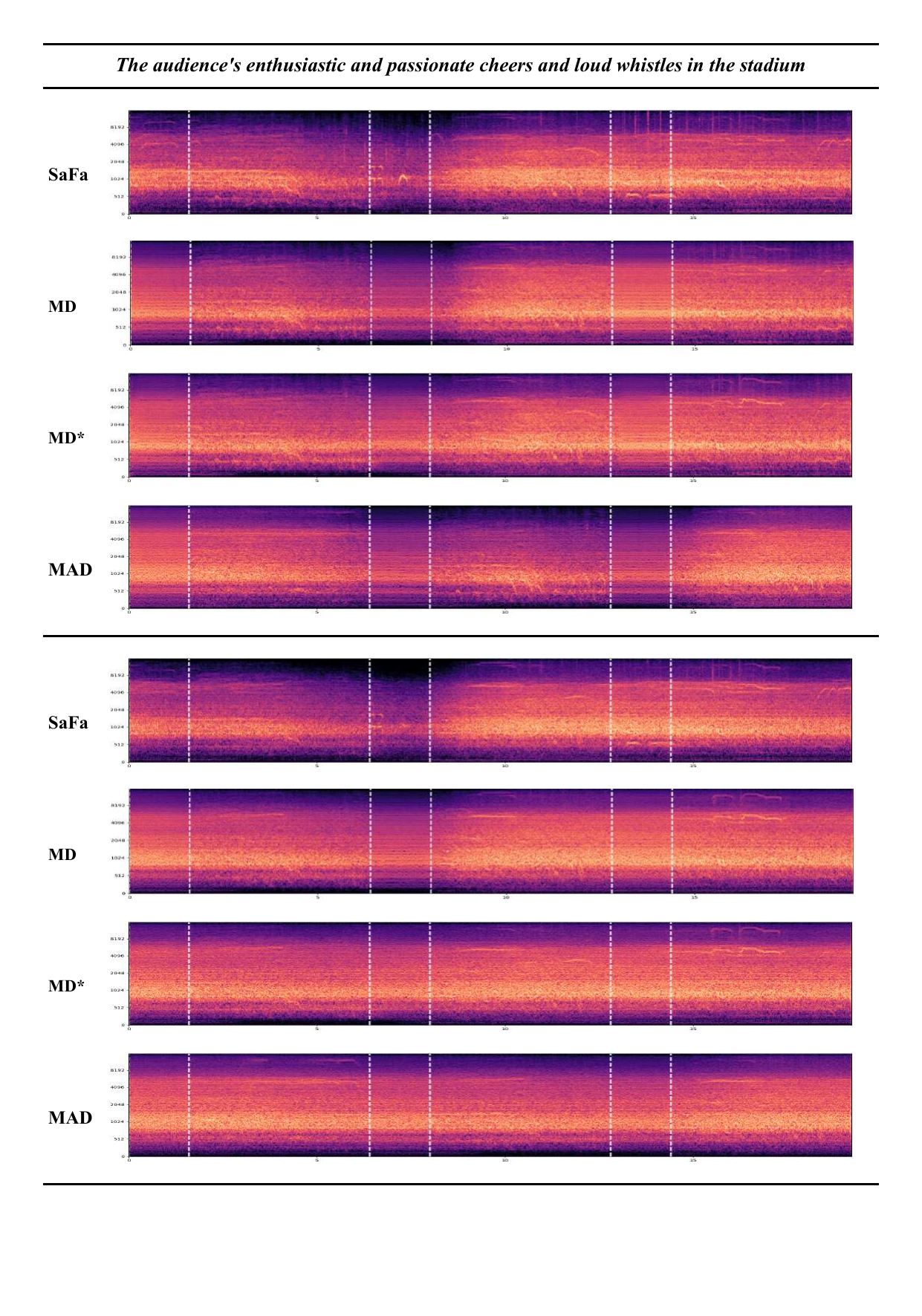}

\vspace{-60pt}
\caption{Qualitative comparison on soundscape generation. MD* represent an enhanced MD method with triangular windows.}
\label{fig:audio_1}
\end{figure*}

\begin{figure*}[t]
    \centering
    \setlength{\belowcaptionskip}{-10pt} 
    \includegraphics[width=2.0\columnwidth]{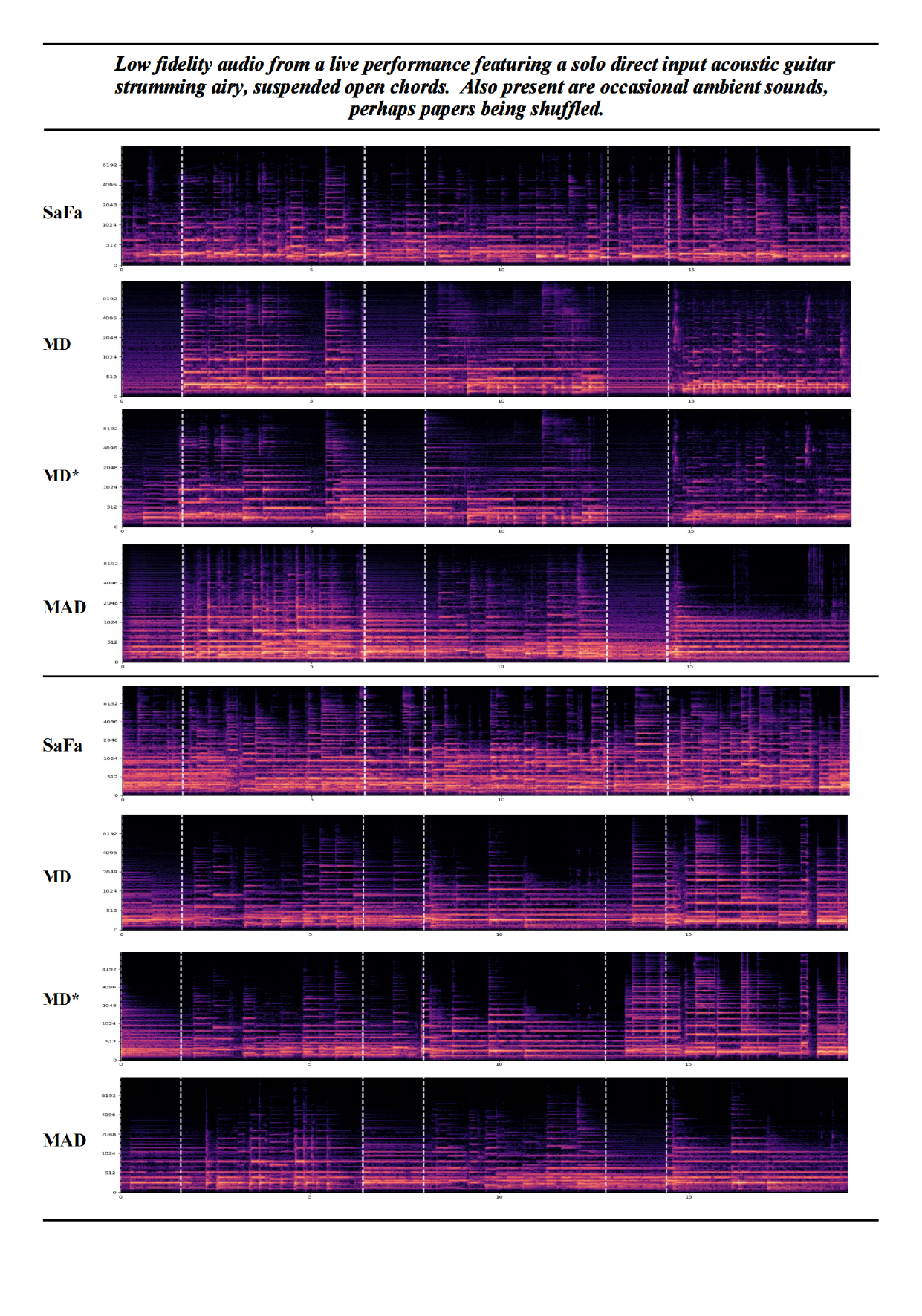}

\vspace{-50pt}
\caption{Qualitative comparison on music generation. MD* represent an enhanced MD method with triangular windows.}
\label{fig:soft_music_1}
\end{figure*}

\begin{figure*}[t]
    \centering
    \setlength{\belowcaptionskip}{-10pt} 
    \includegraphics[width=2.0\columnwidth]{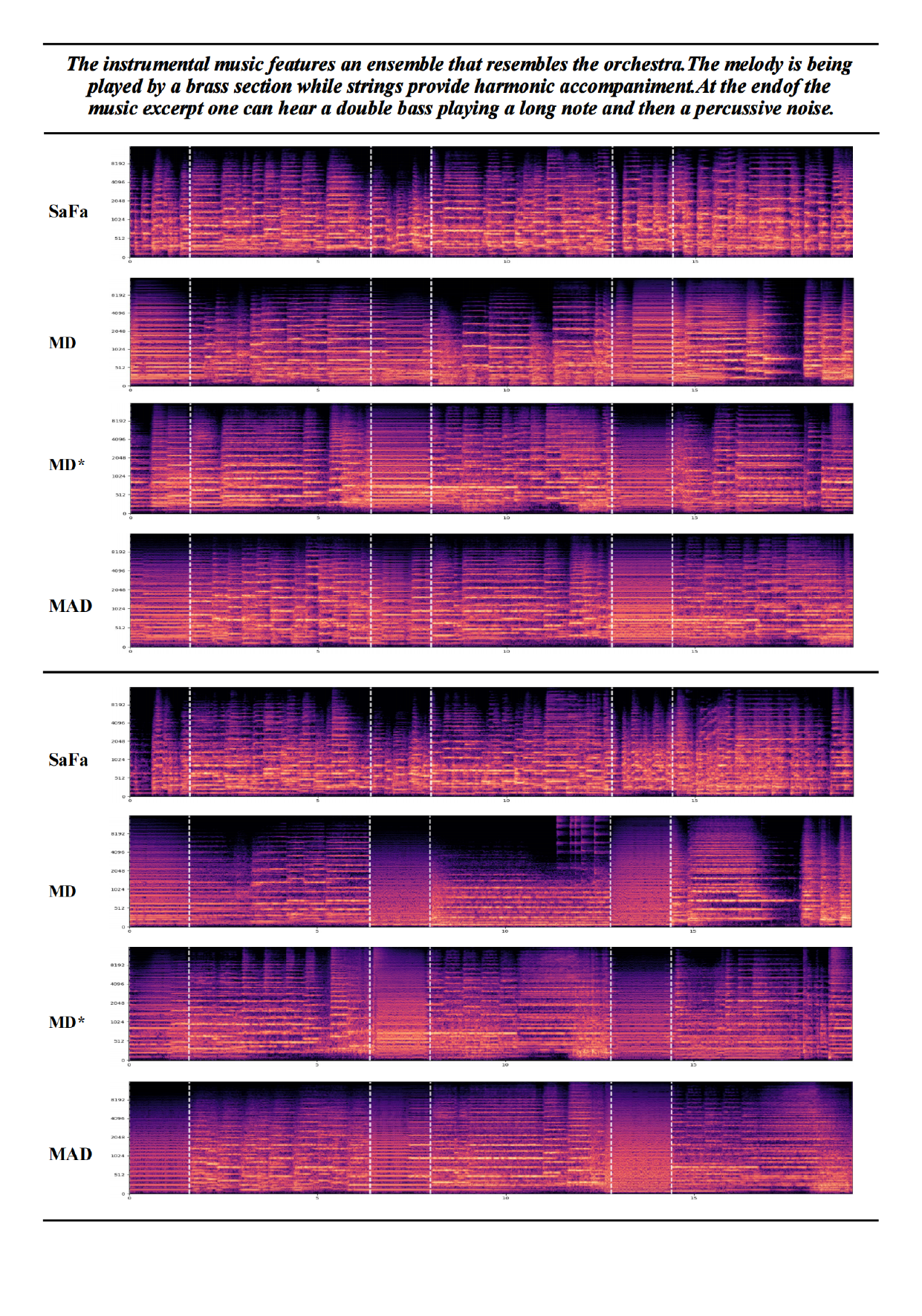}

\vspace{-50pt}
\caption{Qualitative comparison on music generation. MD* represent an enhanced MD method with triangular windows.}
\label{fig:soft_music_2}
\end{figure*}

\begin{figure*}[t]
    \centering
    \setlength{\belowcaptionskip}{-10pt} 
    \includegraphics[width=1.8\columnwidth]{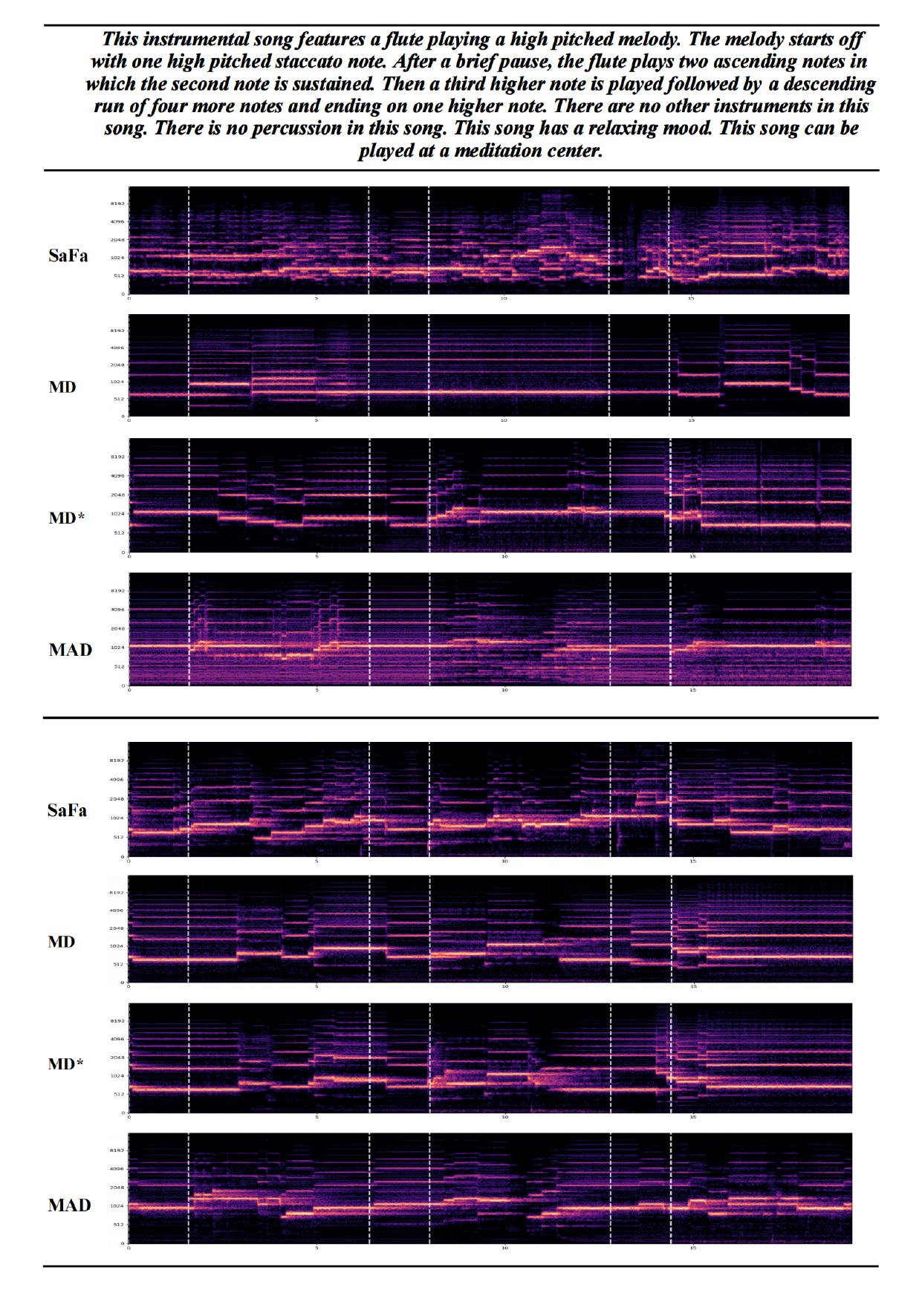}

\vspace{-10pt}
\caption{Qualitative comparison on music generation. MD* represent an enhanced MD method with triangular windows.}
\label{fig:soft_music_3}
\end{figure*}

\begin{figure*}[t]
    \centering
    \setlength{\belowcaptionskip}{-10pt} 
    \includegraphics[width=2.0\columnwidth]{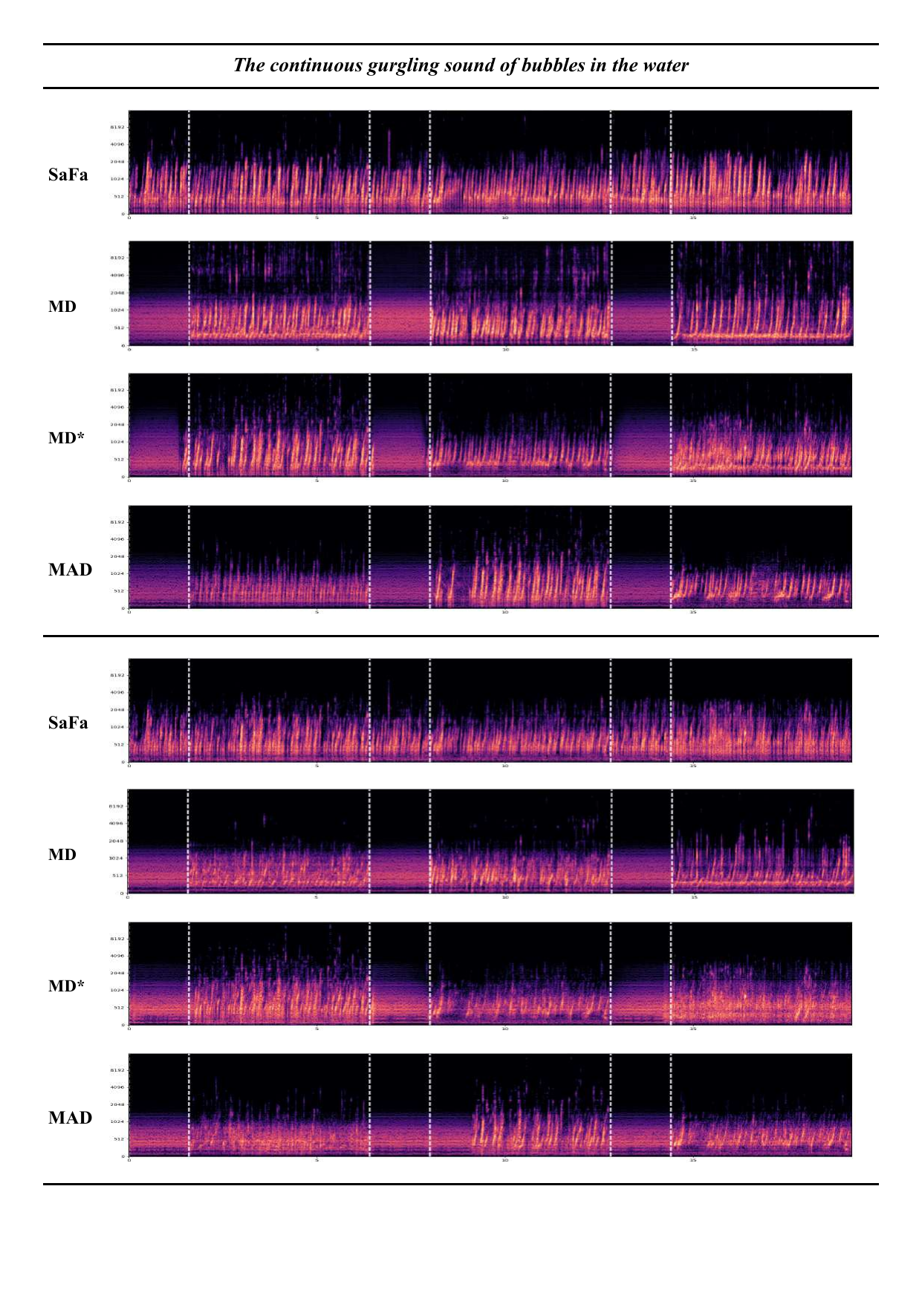}

\vspace{-60pt}
\caption{Qualitative comparison on audio effect generation. MD* represent an enhanced MD method with triangular windows.}
\label{fig:audio_4}
\end{figure*}

\begin{figure*}[t]
    \centering
    \setlength{\belowcaptionskip}{-10pt} 
    \includegraphics[width=2.0\columnwidth]{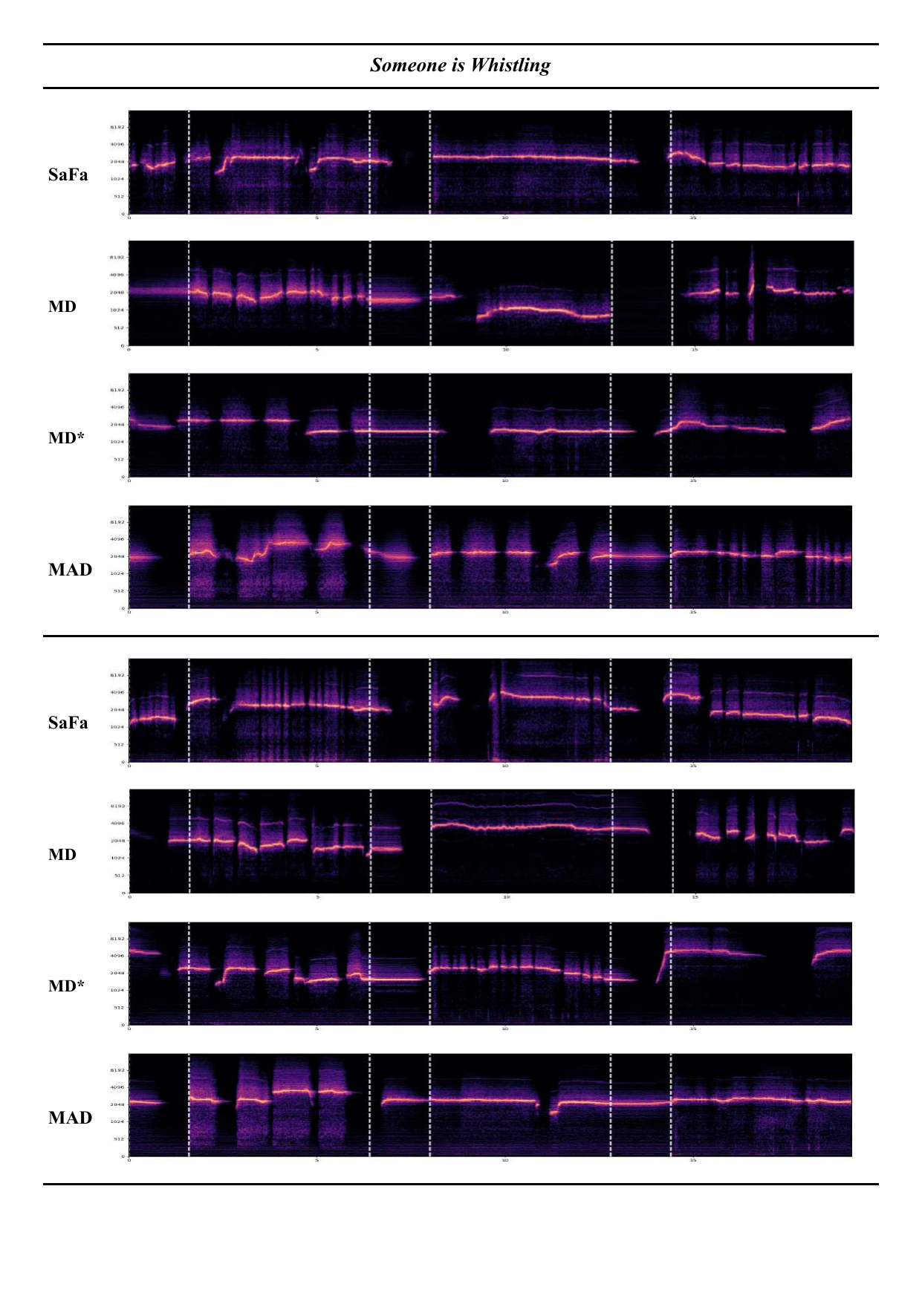}
    
\vspace{-60pt}
\caption{Qualitative comparison on audio effect generation. MD* represent an enhanced MD method with triangular windows.}
\label{fig:audio_2}
\end{figure*}

\begin{figure*}[t]
    \centering
    \setlength{\belowcaptionskip}{-10pt} 
    \includegraphics[width=2.0\columnwidth]{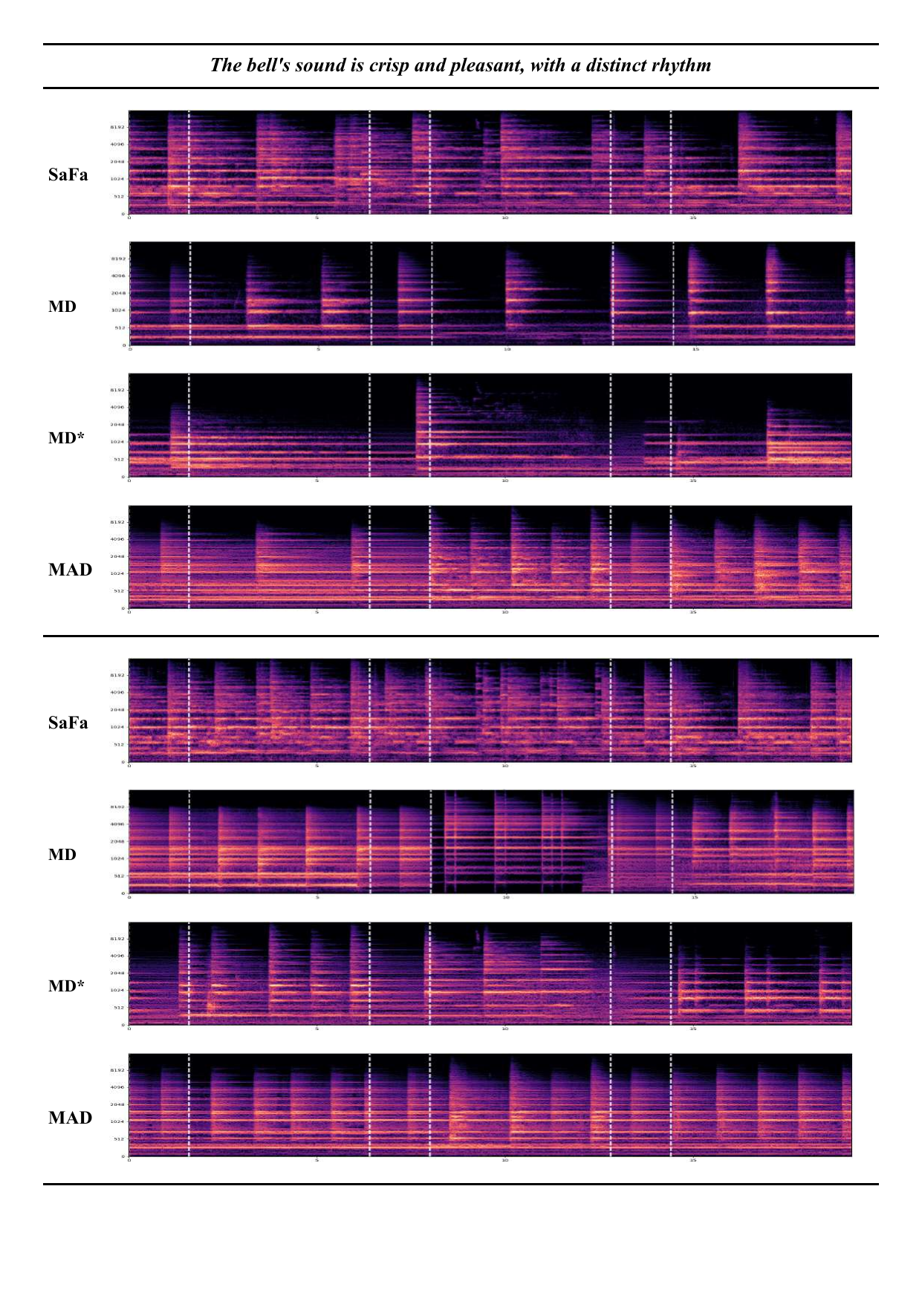}
\vspace{-60pt}
\caption{Qualitative comparison on audio effect generation. MD* represent an enhanced MD method with triangular windows.}
\label{fig:audio_3}
\end{figure*}

\begin{figure*}[t]
\centering
\setlength{\belowcaptionskip}{-10pt} 
\includegraphics[width=2.0\columnwidth]{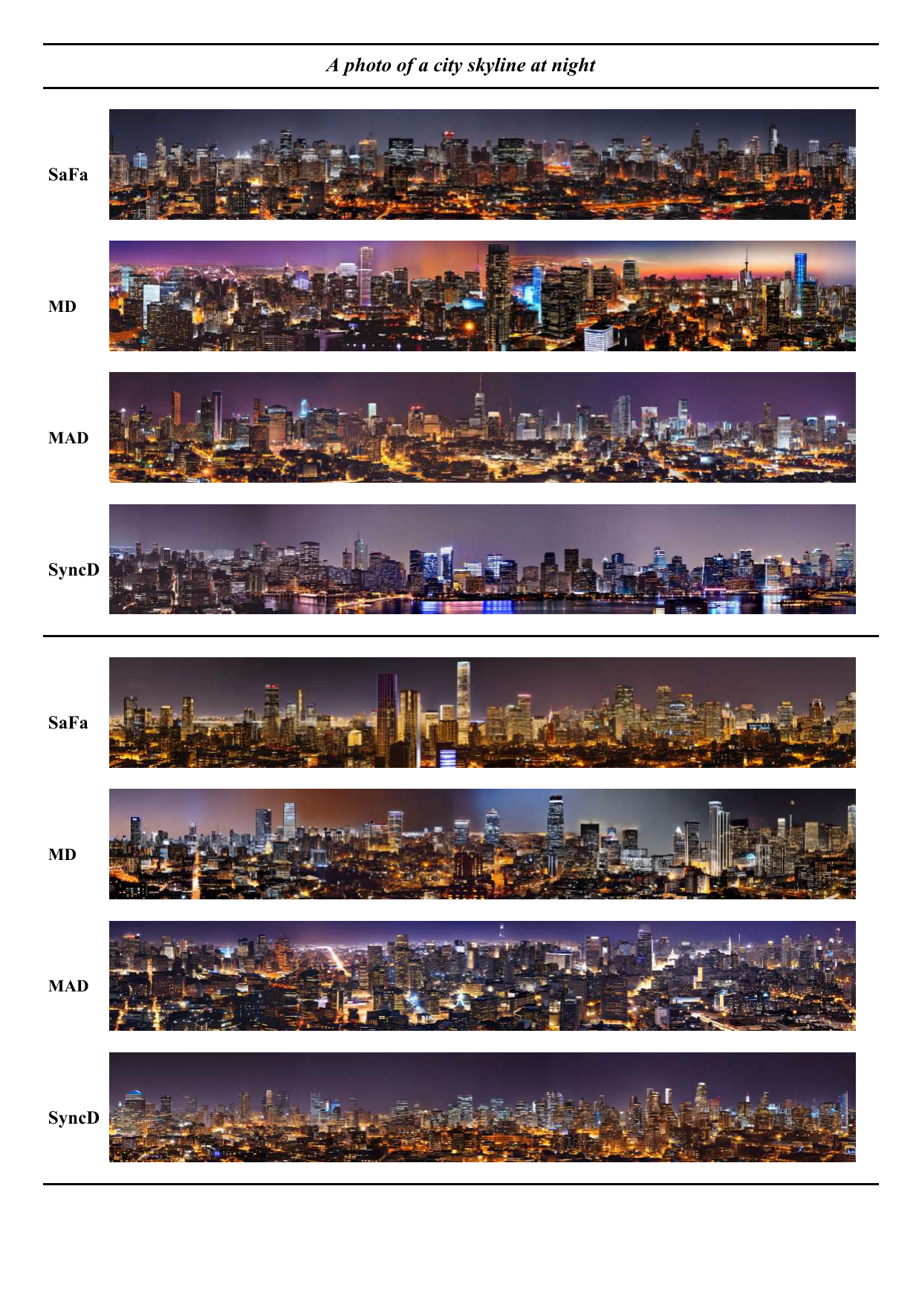}
\vspace{-60pt}
\caption{Qualitative comparison on panorama image generation. MD* represent an enhanced MD method with triangular windows.}
\label{fig:image_0}
\end{figure*}

\begin{figure*}[t]
    \centering
    \setlength{\belowcaptionskip}{-10pt} 
    \includegraphics[width=2.0\columnwidth]{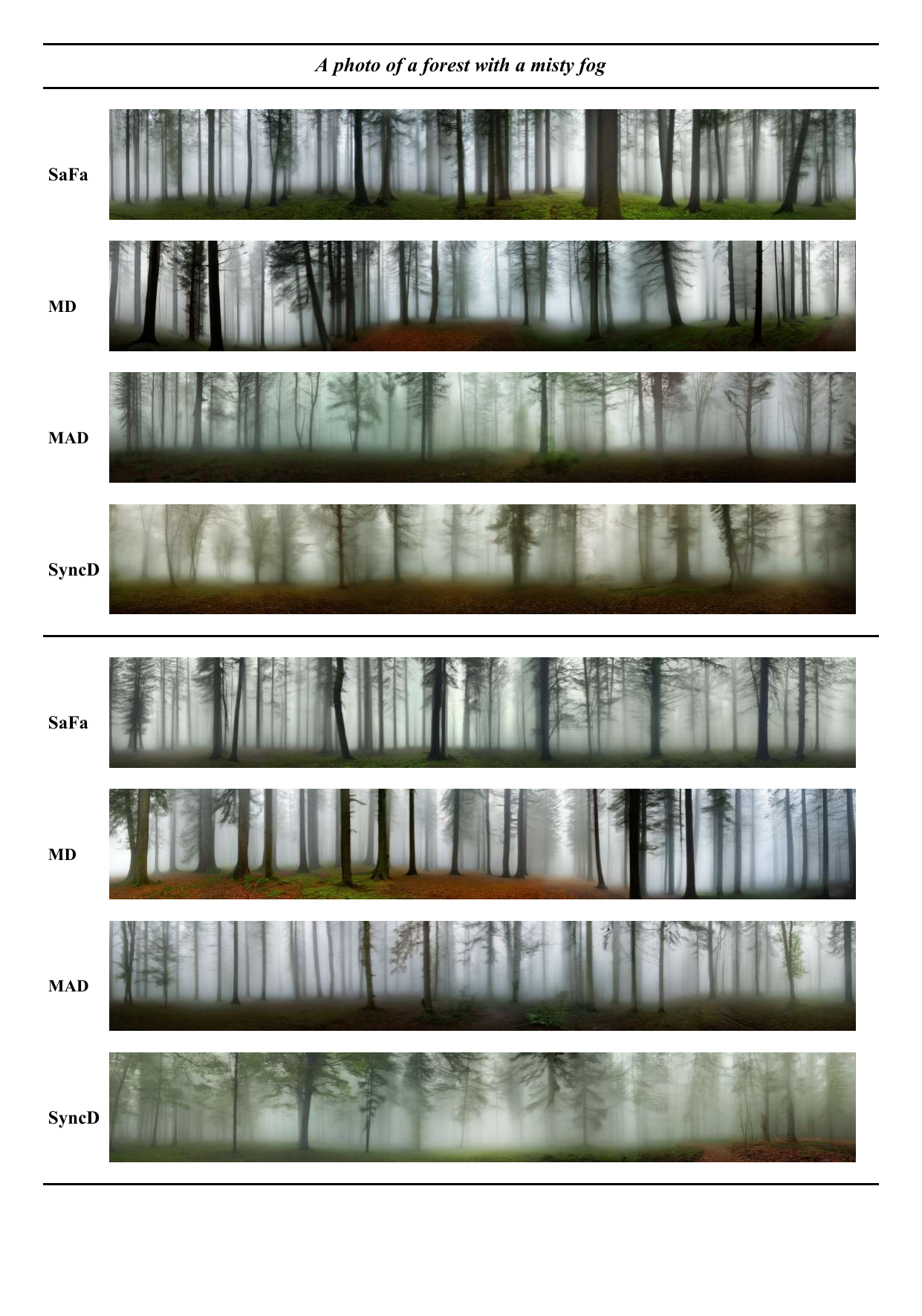}
\vspace{-60pt}
\caption{Qualitative comparison on panorama image generation. MD* represent an enhanced MD method with triangular windows.}
\label{fig:image_1}
\end{figure*}

\begin{figure*}[t]
    \centering
    \setlength{\belowcaptionskip}{-10pt} 
    \includegraphics[width=2.0\columnwidth]{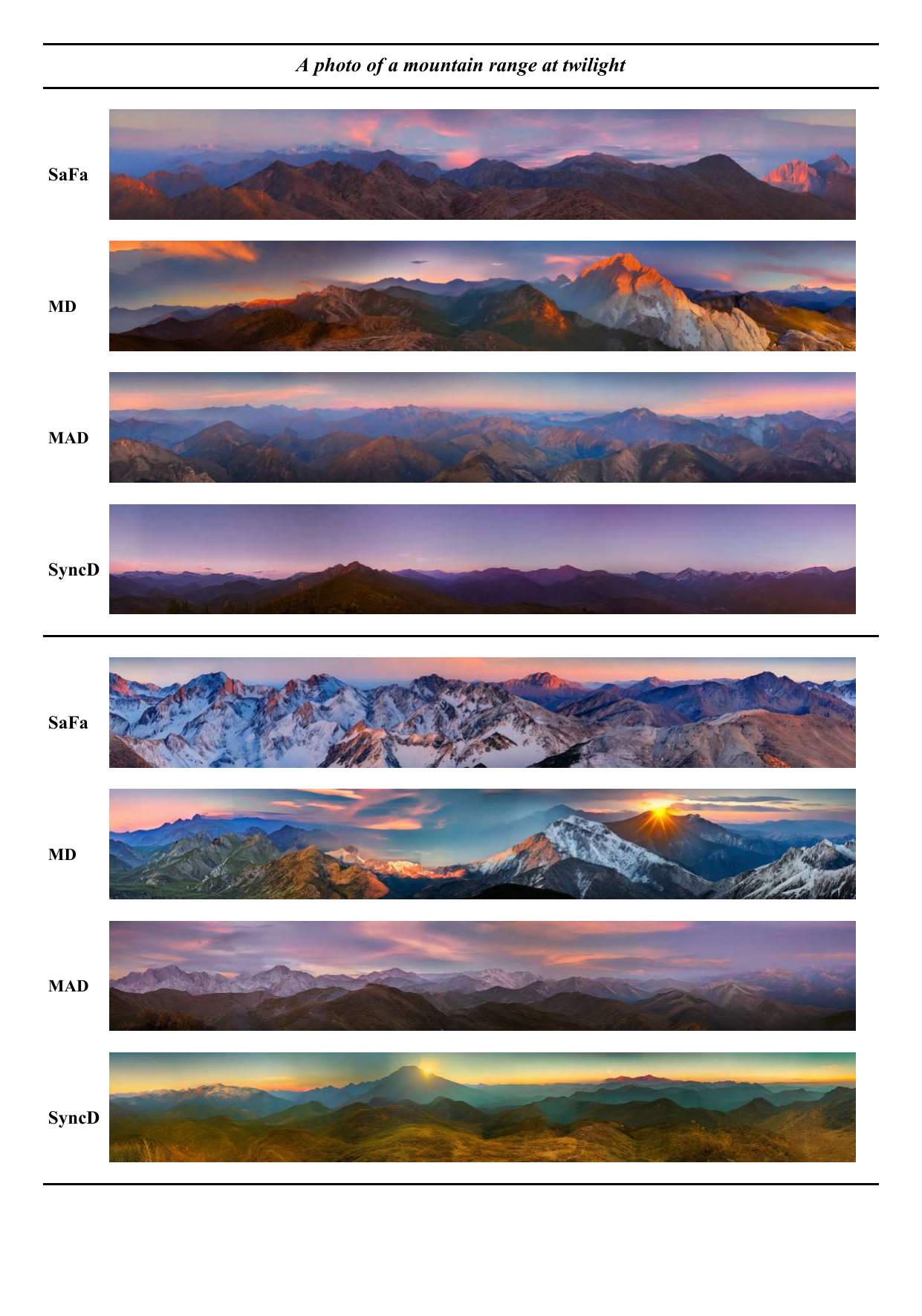}

\vspace{-60pt}
\caption{Qualitative comparison on panorama image generation. MD* represent an enhanced MD method with triangular windows.}
\label{fig:image_2}
\end{figure*}

\begin{figure*}[t]
    \centering
    \setlength{\belowcaptionskip}{-10pt} 
    \includegraphics[width=2.0\columnwidth]{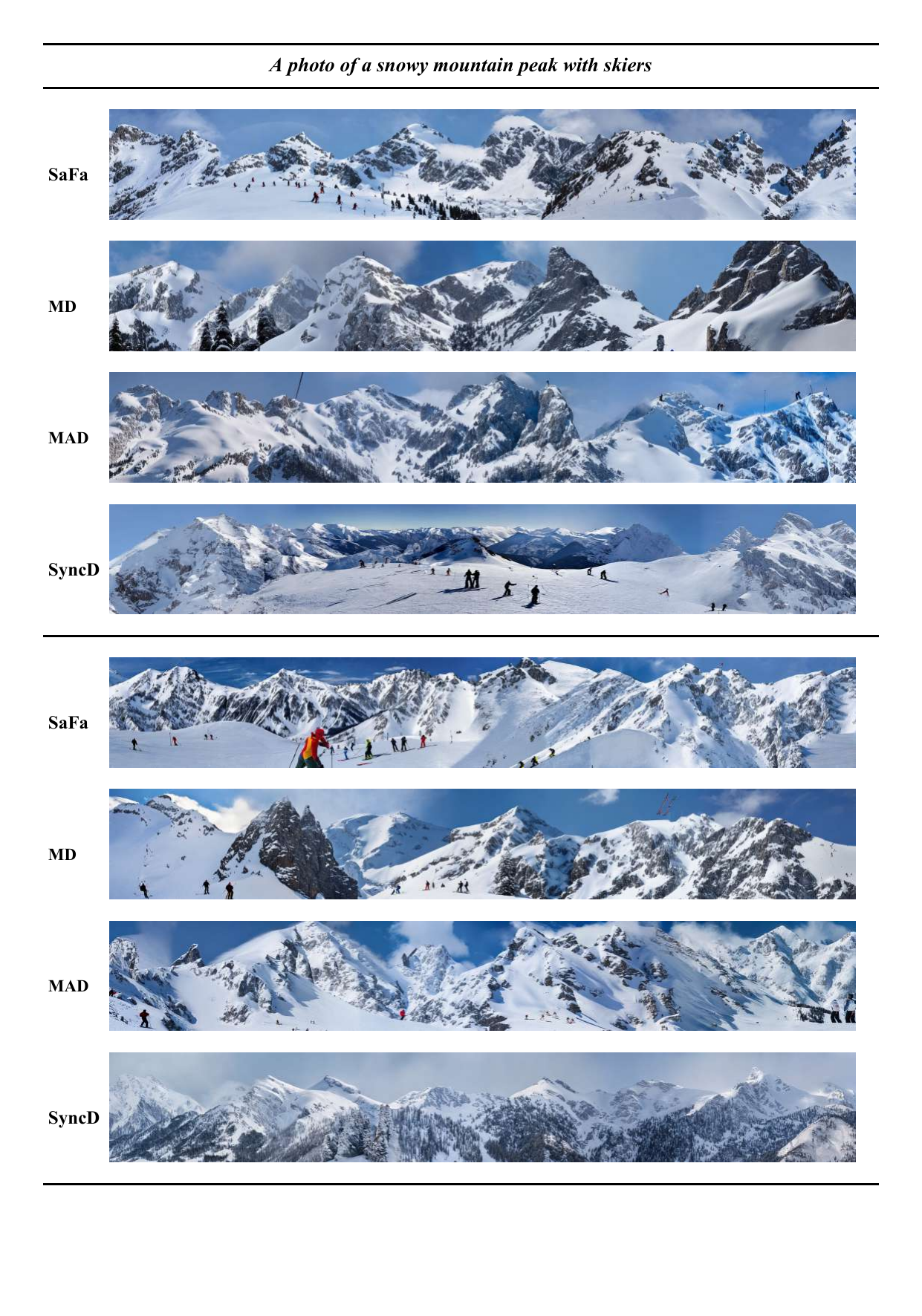}

\vspace{-60pt}
\caption{Qualitative comparison on panorama image generation. MD* represent an enhanced MD method with triangular windows.}
\label{fig:image_3}
\end{figure*}

\begin{figure*}[t]
    \centering
    \setlength{\belowcaptionskip}{-10pt} 
    \includegraphics[width=2.0\columnwidth]{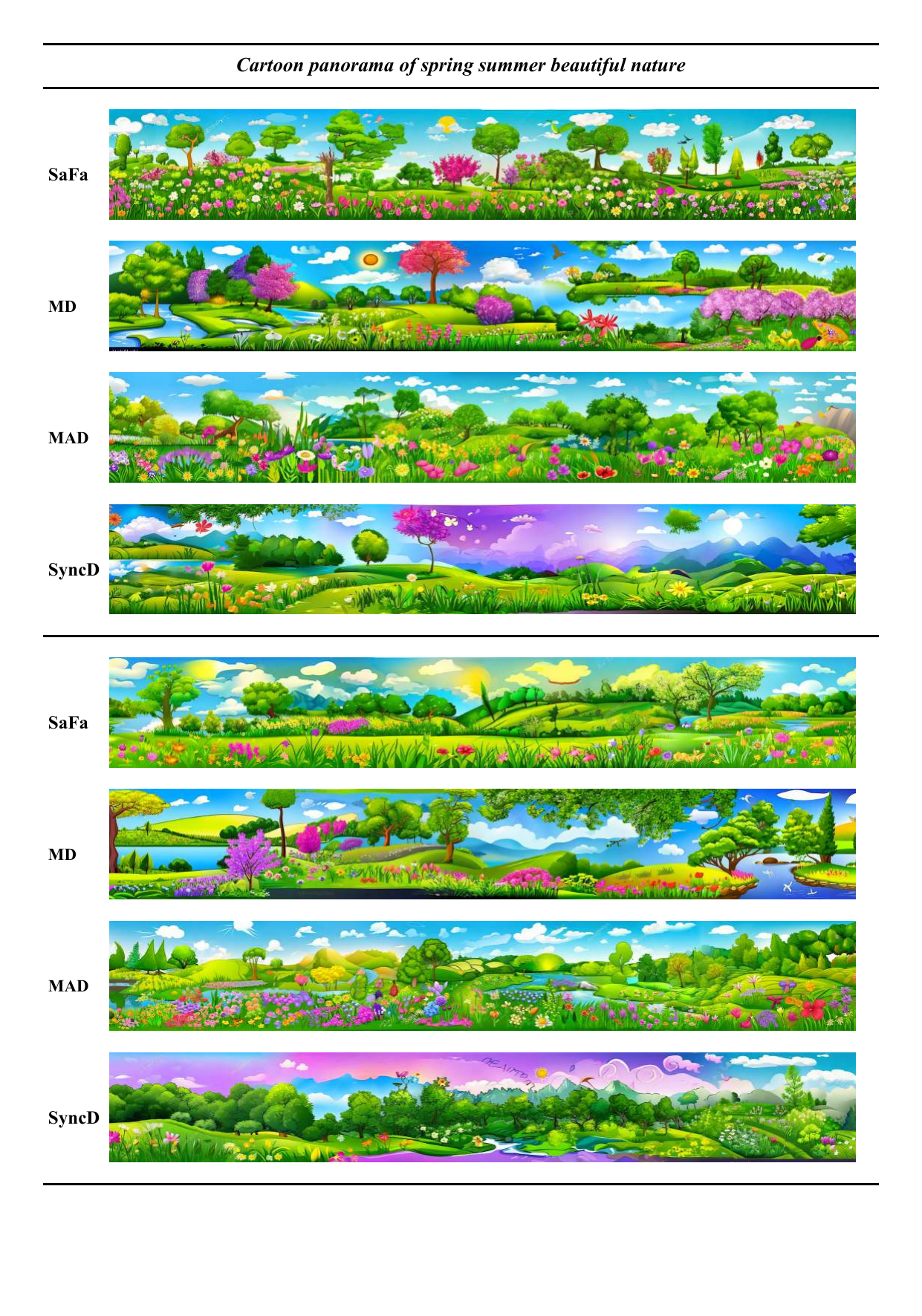}

\vspace{-60pt}
\caption{Qualitative comparison on panorama image generation. MD* represent an enhanced MD method with triangular windows.}
\label{fig:image_4}
\end{figure*}

\begin{figure*}[t]
    \centering
    \setlength{\belowcaptionskip}{-10pt} 
    \includegraphics[width=2.0\columnwidth]{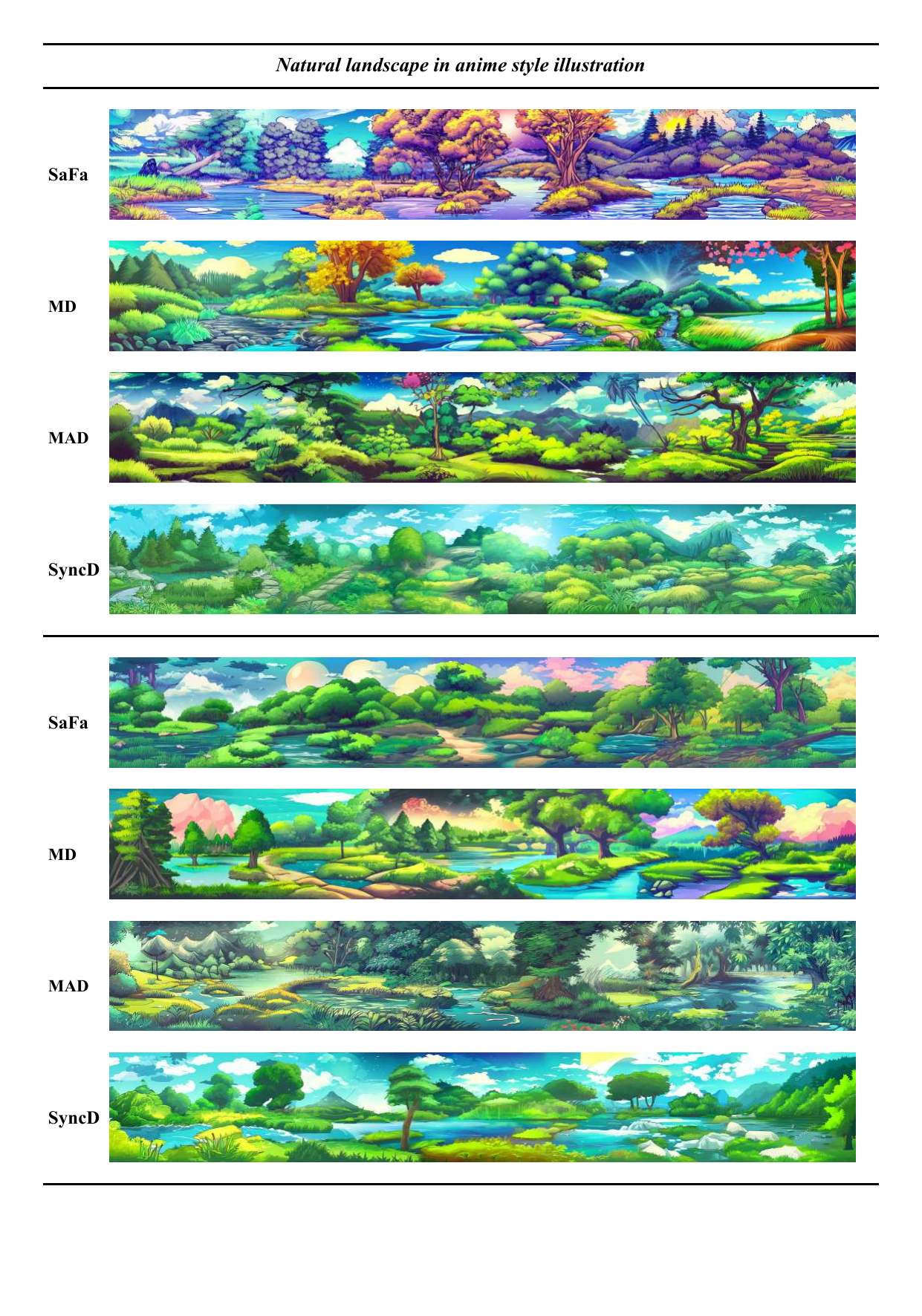}

\vspace{-60pt}
\caption{Qualitative comparison on panorama image generation. MD* represent an enhanced MD method with triangular windows.}
\label{fig:image_5}
\end{figure*}

\begin{figure*}[t]
    \centering
    \setlength{\belowcaptionskip}{-10pt} 
    \includegraphics[width=2.0\columnwidth]{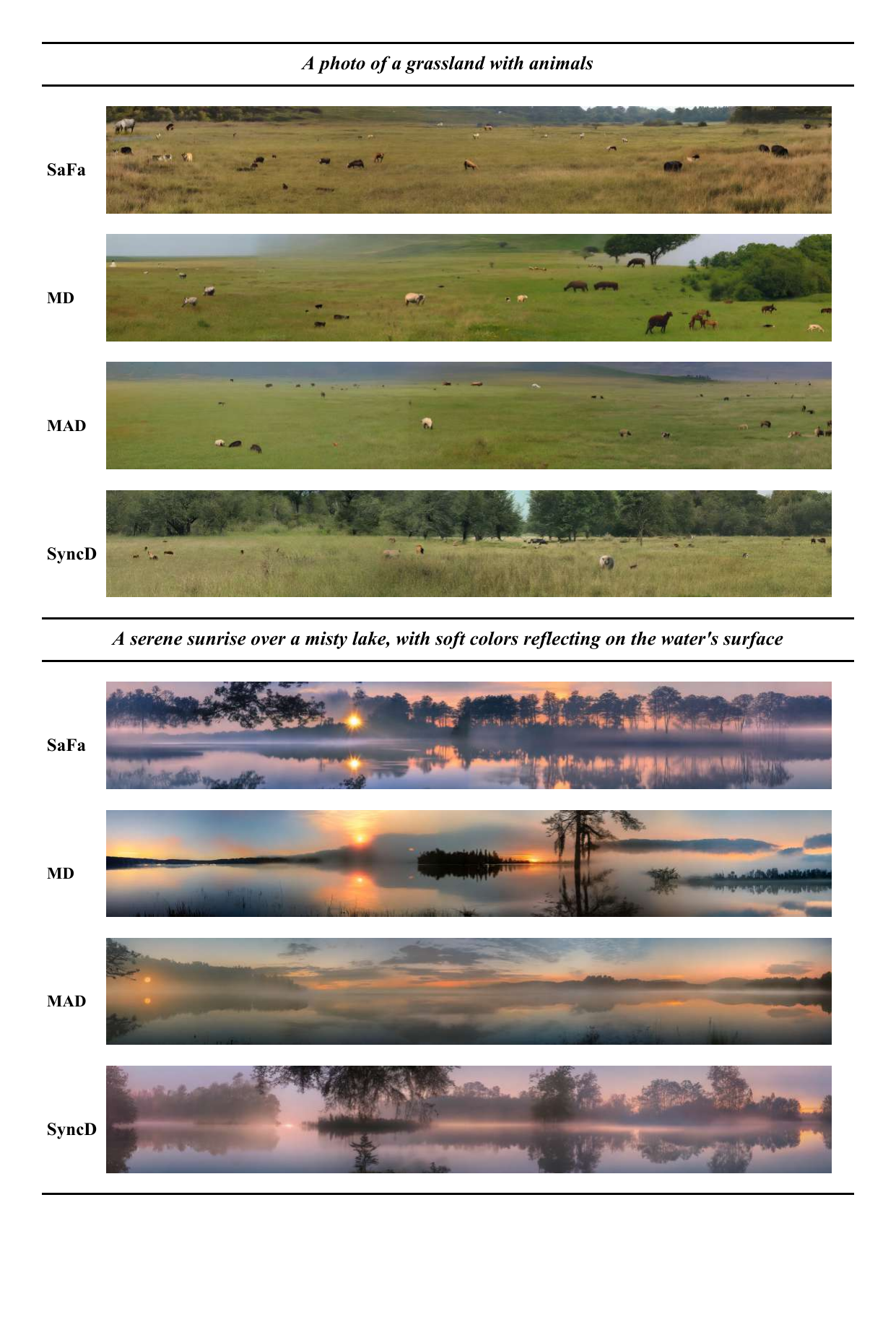}

\vspace{-60pt}
\caption{Qualitative comparison on panorama image generation. MD* represent an enhanced MD method with triangular windows.}
\label{fig:img4_0}
\end{figure*}

\begin{figure*}[t]
    \centering
    \setlength{\belowcaptionskip}{-10pt} 
    \includegraphics[width=2.0\columnwidth]{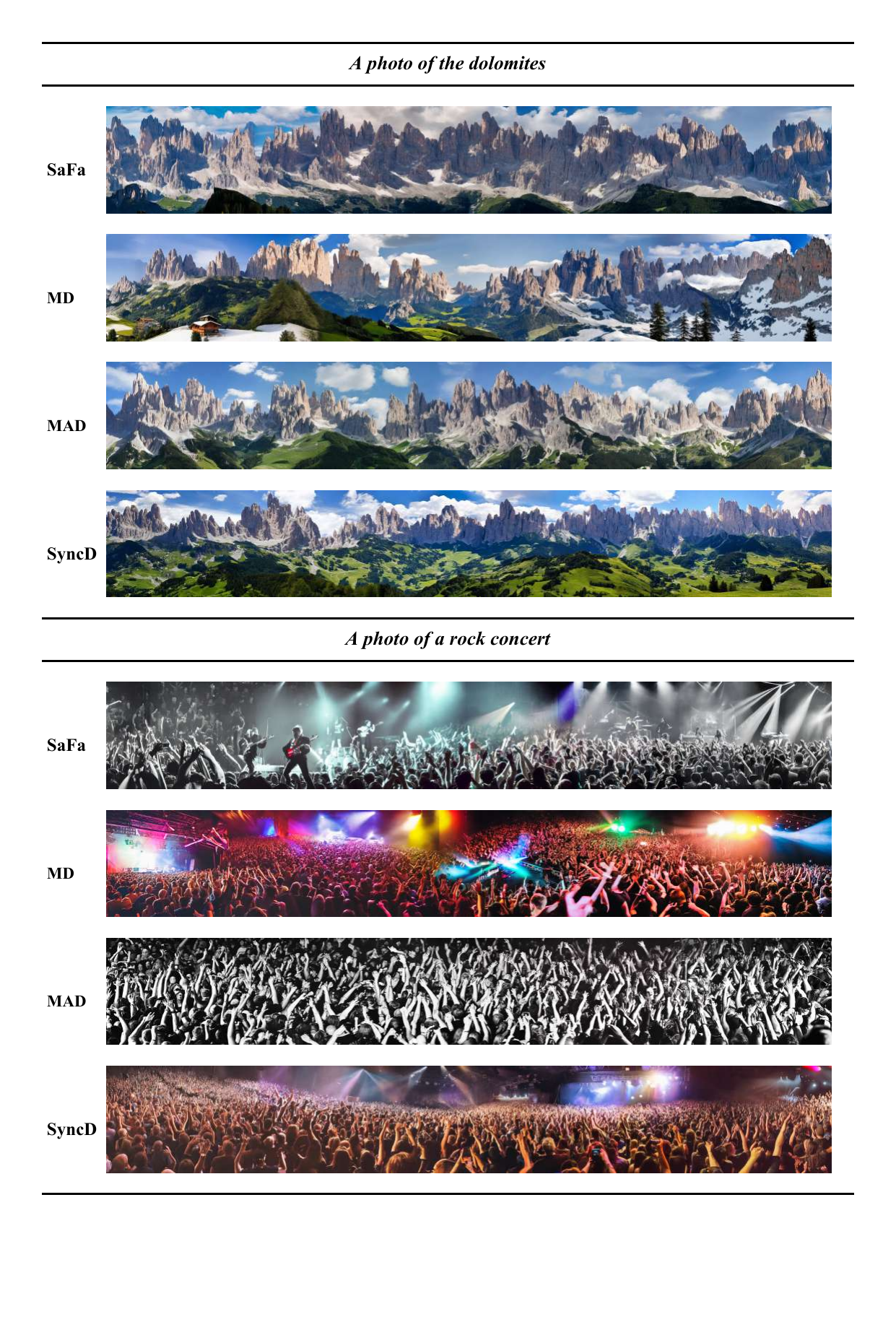}

\vspace{-60pt}
\caption{Qualitative comparison on panorama image generation. MD* represent an enhanced MD method with triangular windows.}
\label{fig:img4_1}
\end{figure*}

\begin{figure*}[t]
    \centering
    \setlength{\belowcaptionskip}{-10pt} 
    \includegraphics[width=2.0\columnwidth]{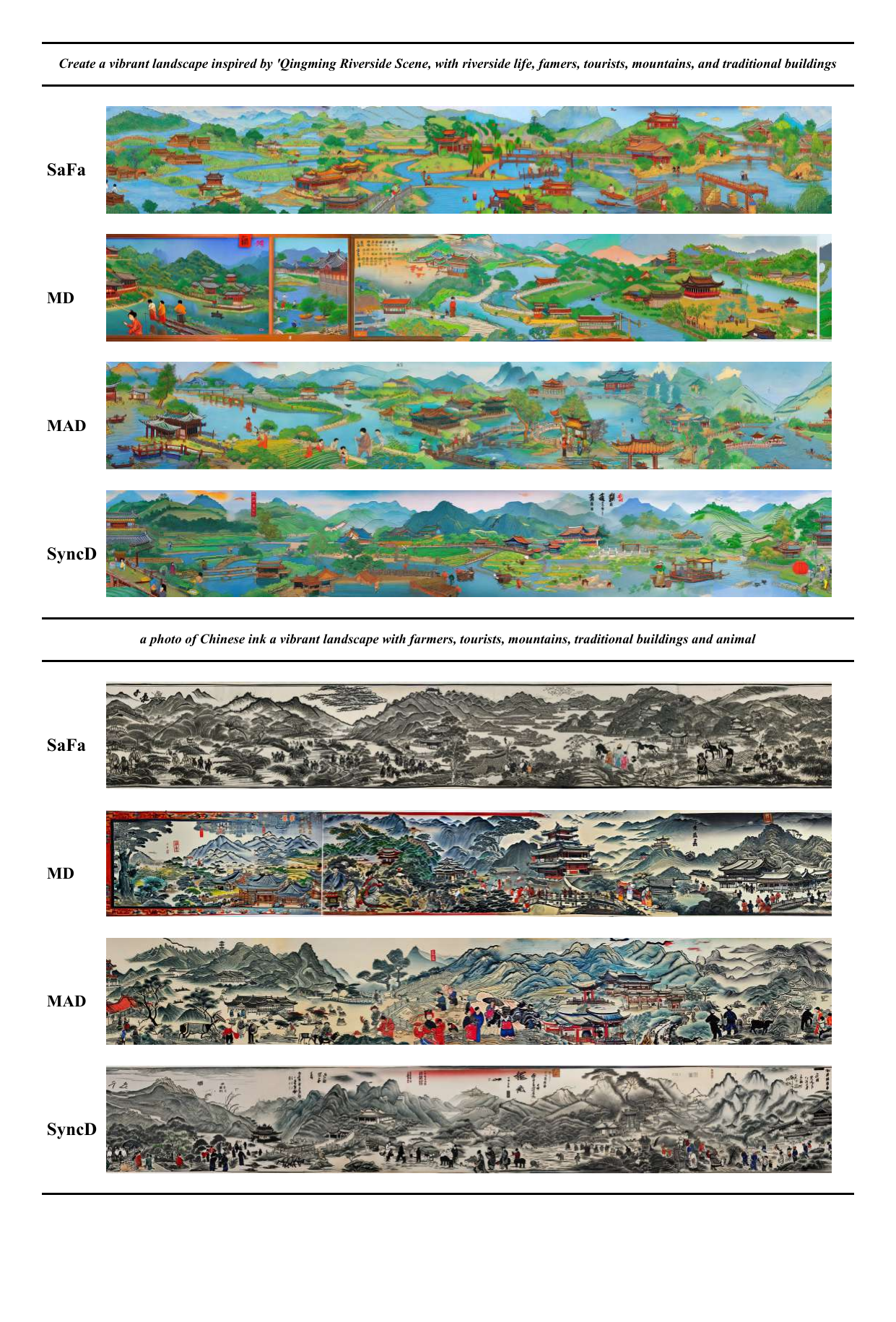}

\vspace{-60pt}
\caption{Qualitative comparison on panorama image generation. MD* represent an enhanced MD method with triangular windows.}
\label{fig:img4_2}
\end{figure*}

\begin{figure*}[t]
    \centering
    \setlength{\belowcaptionskip}{-10pt} 
    \includegraphics[width=2.0\columnwidth]{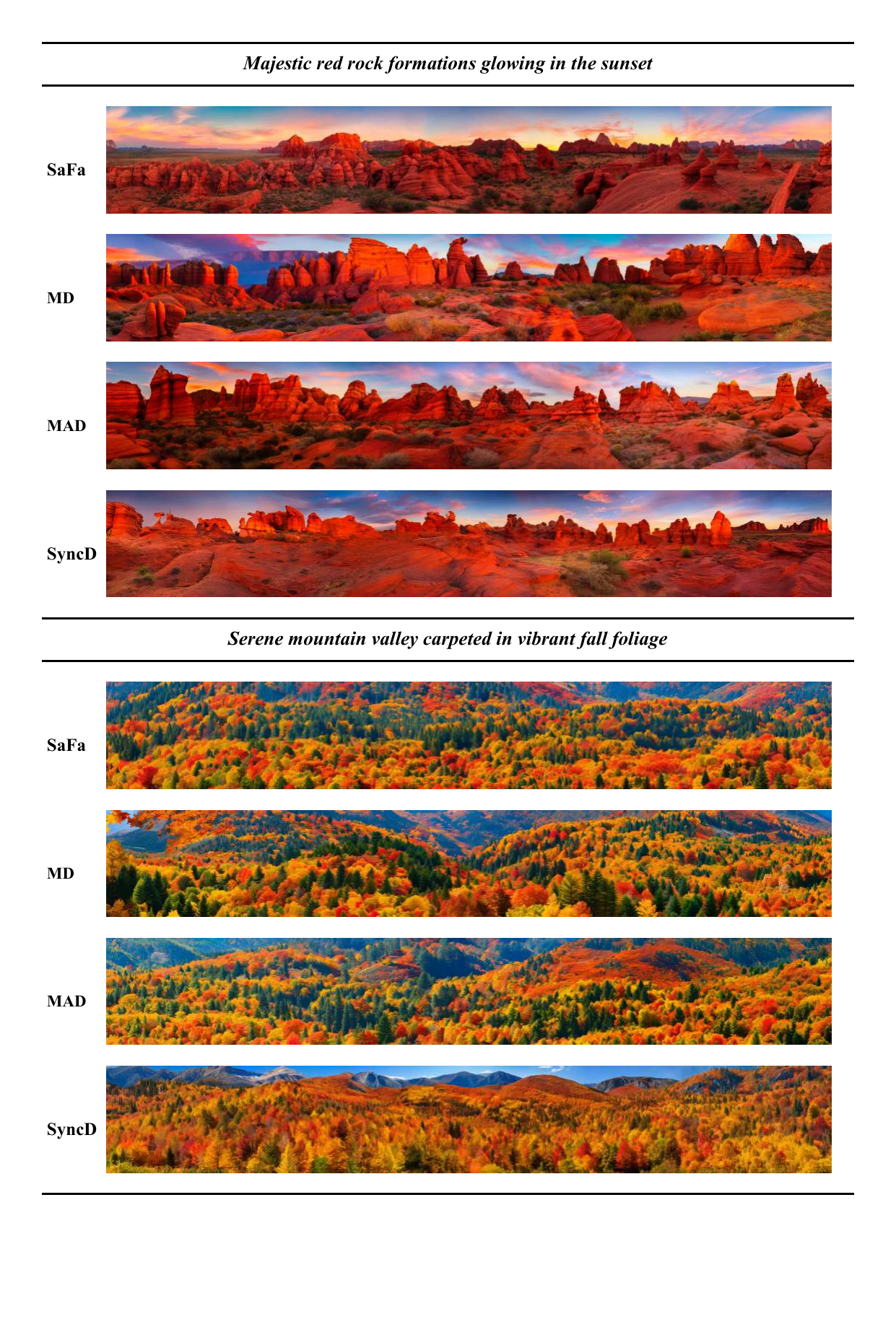}

\vspace{-60pt}
\caption{Qualitative comparison on panorama image generation. MD* represent an enhanced MD method with triangular windows.}
\label{fig:img4_3}
\end{figure*}

\begin{figure*}[t]
    \centering
    \setlength{\belowcaptionskip}{-10pt} 
    \includegraphics[width=2.0\columnwidth]{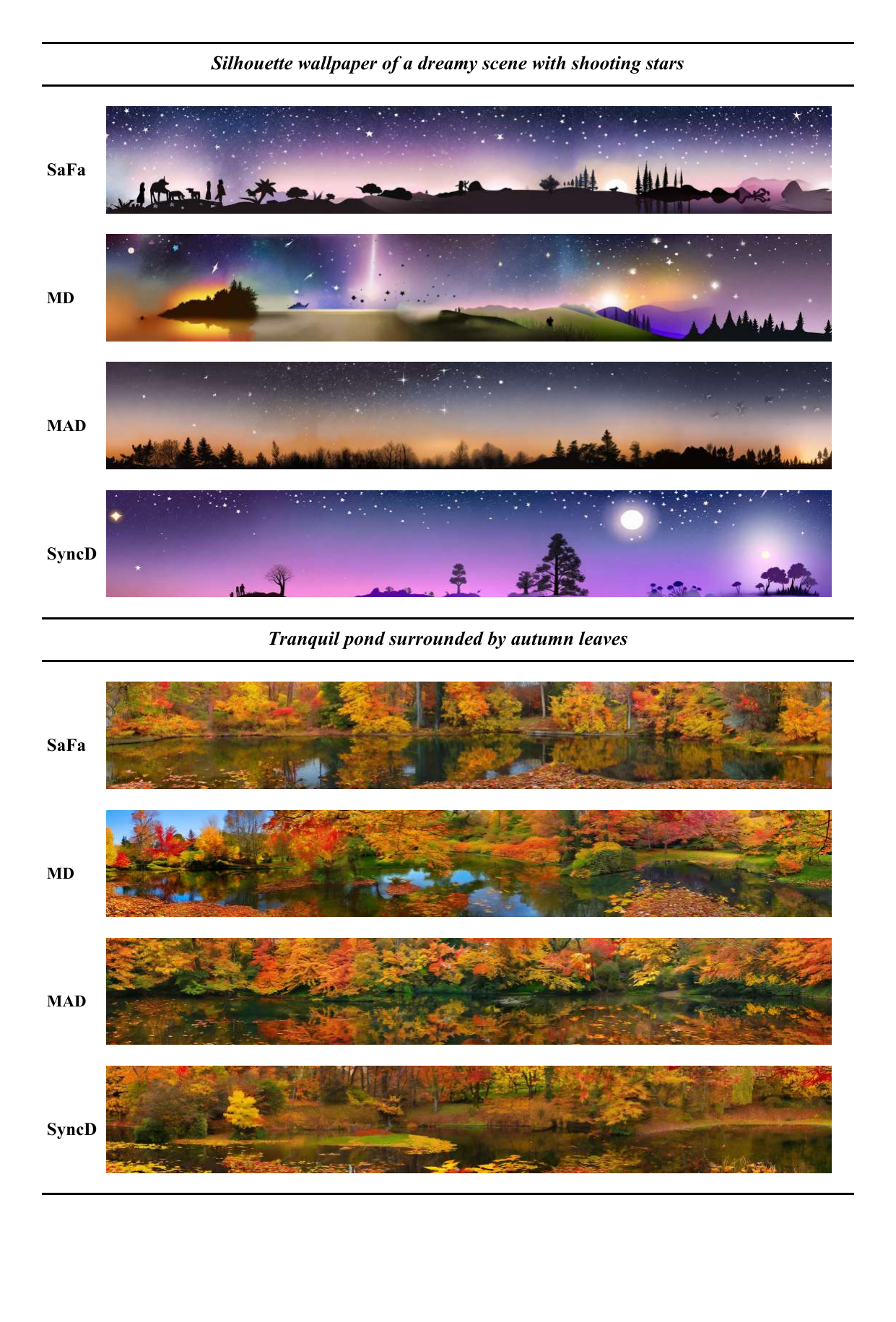}

\vspace{-60pt}
\caption{Qualitative comparison on panorama image generation. MD* represent an enhanced MD method with triangular windows.}
\label{fig:img4_5}
\end{figure*}

More qualitative results on the audio generation are in Fig. \ref{fig:audio_0} to \ref{fig:audio_3} and panorama generation are in Fig. \ref{fig:image_0} to \ref{fig:img4_5}.

\twocolumn 

\end{document}